\documentclass[aip, rsi,reprint,10pt]{revtex4-1}
\usepackage{amsmath}
\usepackage{graphicx}
\usepackage{verbatim}
\usepackage{color}
\usepackage{subfigure}
\usepackage{hyperref}

\begin{document}

\title{A versatile apparatus for two-dimensional atomtronic quantum simulation}
\author{T. A. Haase}
\email{thaa191@aucklanduni.ac.nz}
\author{D. H. White}
\altaffiliation{Present address: Department of Applied Physics, Waseda University, 3-4-1 Okubo, Shinjuku-ku, Tokyo 169-8555, Japan.}
\author{D. J. Brown}
\author{I. Herrera}
\author{M. D. Hoogerland}
\affiliation{Dodd-Walls Centre for Photonic \& Quantum Technologies, Department of Physics, University of Auckland, Private Bag 92019, Auckland, New Zealand}

\begin{abstract}
We report on the implementation of a novel optical setup for generating high-resolution customizable potentials to address ultracold bosonic atoms in two dimensions. Two key features are developed for this purpose. The customizable potential is produced with a direct image of a spatial light modulator, conducted with an in-vacuum imaging system of high numerical aperture. Custom potentials are drawn over an area of $600\times 400~\mu$m with a resolution of 0.9~$\mu$m. The second development is a two-dimensional planar trap for atoms with an aspect ratio of 900 and spatial extent of Rayleigh range 1.6~$\times$~1.6~mm, providing near-ballistic in-planar movement. We characterize the setup and present a brief catalog of experiments to highlight the versatility of the system.

\end{abstract}

\maketitle

\section{Introduction}

Quantum simulation is an active area of research in which the behavior of complex quantum systems is modeled by simpler, more controllable physical systems~\cite{Bloch2012,Omalley2016,Cirac2012,Lanyon2011,Sherson2010}. For instance, the wave propagation of electrons in a solid, which determines the electrical properties of materials, can be modeled by the wave propagation of ultracold atoms sourced from a Bose-Einstein Condensate (BEC)~\cite{Chien2015,Billy2008}. Such atoms are cold enough that their thermal de Broglie wavelength is larger than the wavelength of visible light, allowing optical structures to be created with similar dimension to the thermal wavelength. This allows for the dynamics of complex systems to be investigated by studying the motion of ultracold atoms in an analogous optical potential. The high degree of controllability inherent in ultracold atom systems means that the dynamical effects of additional layers of complexity can be systematically studied. 

These potentials to manipulate neutral atoms are commonly produced by the off-resonant AC Stark shift, which is proportional to the light intensity. Gradients in the light intensity hence generate a force, referred to as the `dipole force'. Arbitrary potentials can be generated by spatial light modulators (SLMs)\cite{Greiner_DMD_phase}, digital mirror devices (DMDs) \cite{Gauthier,optical_tweezers_atoms,Liang:09}, or by `painting' time-averaged potentials \cite{painted_potentials,PaintedRing}. In this way, the `programming' on such quantum simulators is performed through the creation of customizable optical potentials to create a Hamiltonian engineered to mimic the desired system. 

Application of these light-shaping methods has allowed for the control of cold atoms on varying length scales. On one hand, careful preparation of single atoms on sub-micron lattices has been achieved~\cite{Greiner_Atom_Microscope}. On the other hand, novel trap geometries have led to the advent of atomtronics, where atomic analogues of electronic circuits are created over length scales of tens of micrometers. Geometries such as ring traps have been successful in modeling atomic SQUID circuits~\cite{ToroidalSuperflow,StirredRing,ResistiveFlowRing,RingHysterisis,RingAtomtronicInterference,RingAtomtronicSquid,AtomtronicSQUIDOptimization}, and further trap geometries have been proposed~\cite{AtomtronicCircuits,AtomtronicsAnalogs,AtomtronicTransistor,AtomtronicTransistorPrinciple,ElectricAnalogsAtoms} and realized~\cite{WeakLink,NegativeConductivity,Eckle} for other atomtronic circuits. 

Here we describe our system used to manipulate ultracold $^{87}$Rb atoms trapped in two dimensions. We implement a large area 2D trap which prevents motion in the vertical direction while allowing near-ballistic motion within the plane. At the same time, we project customizable light patterns created by an SLM using intensity modulation. The imaging system developed for the SLM is designed to produce optical structures with a resolution of 1~$\mu$m. The SLM is imaged using 532~nm light, which produces a repulsive potential for $^{87}$Rb. The combination of both these elements allows us to investigate transport dynamics at large length scales in two dimensions, which is a region not yet addressed by existing ultracold atom experiments. The ultimate goal of this setup is to move towards a quantum simulator for transport experiments in 2D disordered ultracold atom systems \cite{50_years_anderson}. 
 
\section{Customizable Potentials for Atoms}

We create customizable potentials by shaping the light field of a 532 nm laser using a Holoeye LC-R 720 SLM. The display is a reflective liquid crystal on silicon (LCoS)~\cite{liquid_crystal} chip with a resolution of 1280 $\times{}$ 768 pixels. The pixel pitch is 20~$\mu{}$m and the pixels have 8-bit grayscale resolution. We modulate the light field in intensity, which we find yields images superior to those produced by phase modulation holography~\cite{GS_algorithm,mraf,omraf}. Each pixel of the SLM acts as an adjustable wave plate, which translates to an adjustable intensity through the use of a polariser. The image quality is determined by the quality of the incident beam and induced aberrations from optics. It has been reported that LCOS devices may suffer from intensity flicker~\cite{Garcia-Marquez:12}. We tested this effect on our apparatus by modulating the pixel intensity from off-to-on, and measuring the output with a fast photodiode. We did not detect any significant intensity flicker, and therefore we do not expect significant heating due to laser power fluctuation. 

For our experiments we desire an imaging system capable of producing optical structures of comparable size to the atomic de Broglie wavelength, which is on the order of 1~$\mu$m. This will enable a range of experiments, including exotic lattices~\cite{Kagome, honeycomb}, and will tie in with our future aim of studying disordered systems in 2D.

\begin{figure}[ht]
\centering
\includegraphics[width = 240px, keepaspectratio]{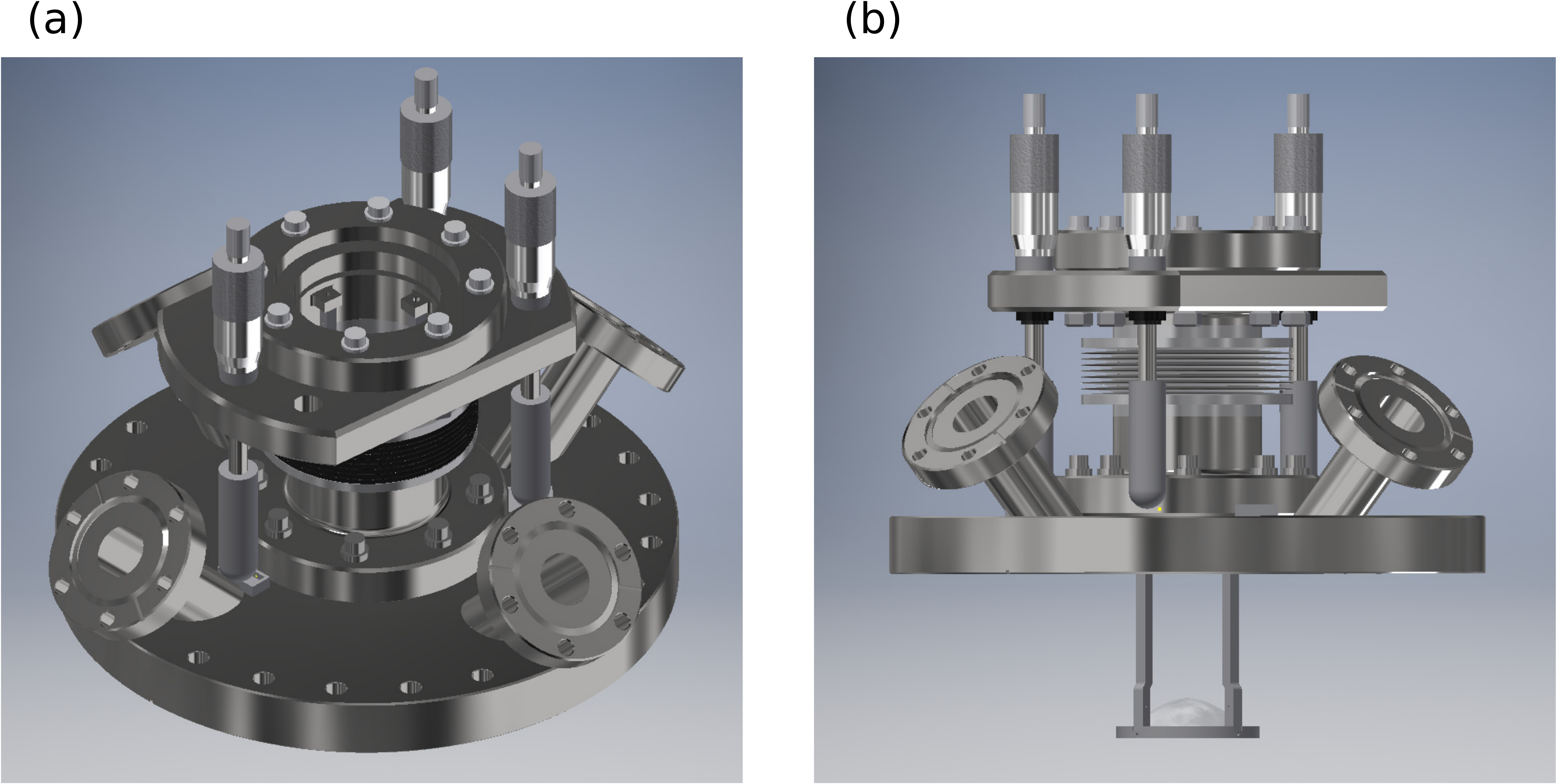}
\caption{(a) CAD view of the complete assembly connected to the top 10" flange of the main vacuum chamber. (b) Side view of the CAD assembly showing the lens mount and mounting rods connecting to the bellows. The rods are designed such that when the bellows are at half of the maximum compression, the lens is in focus with the center of the vacuum chamber. }
\label{fig:bellows}
\end{figure}

To obtain the resolution required, we position a high numerical aperture (NA) lens inside the main vacuum chamber. The location of the magneto optical trap (MOT) beams place an upper bound on the numerical aperture of the imaging lens. We use an aspheric lens with a NA of 0.42 (Edmund Optics part number 67-278) as the final imaging lens inside the chamber. For 532~nm light, we are diffraction limited to a spot size of 800~nm in diameter. We designed a lens mount to hold the asphere in place inside the vacuum chamber, primarily under gravity, with minimal surface contact. The lens mount is connected to a custom vacuum flange which in turn is attached to edge welded bellows, as shown in Fig.~\ref{fig:bellows}. The bellows may be adjusted to control the vertical position and tilt of the the in-vacuum lens. The bellows provide a 20 mm travel length to allow us to position the final image in focus with the cold atom cloud. The in-vacuum lens~\cite{Greiner_Atom_Microscope,Robens:17,Leung:14} allows us to obtain resolutions which are inaccessible with an external lens. The alternative of using a longer working distance lens would reduce the resolution of our system significantly due to the dimensions of our vacuum chamber. An additional advantage of an in-vacuum lens is that the light is collimated at the viewport entrance to the vacuum chamber, which reduces aberrations and shifts induced by the vacuum window.  

\begin{figure}[ht]
\centering
\includegraphics[width = 250px,keepaspectratio]{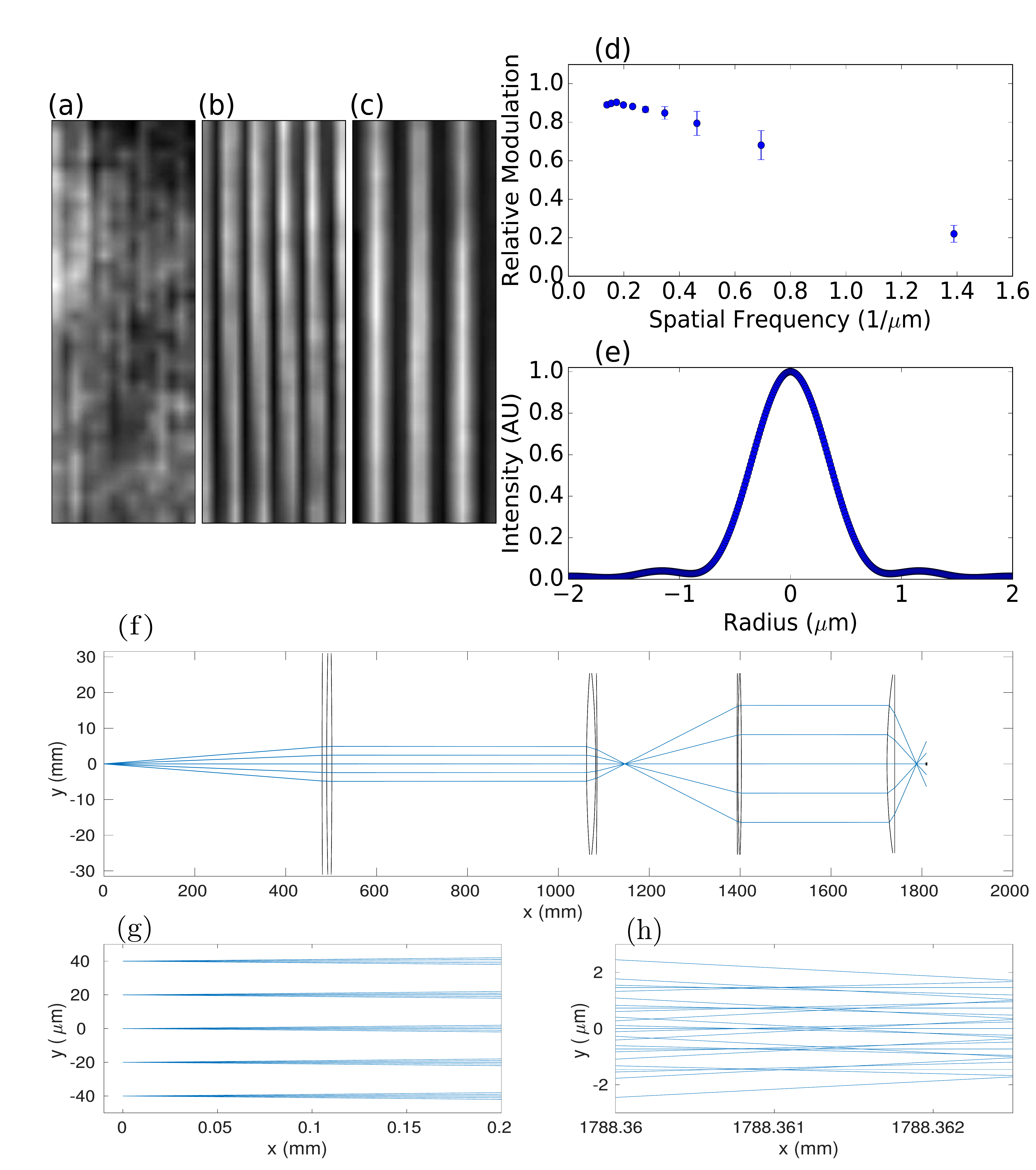}
\caption{Grating images formed by the spatial light modulator are imaged onto a CCD camera using a high numerical aperture (40$\times$) microscope objective. Images of the gratings are shown for: (a) 1~pixel pitch; (b) 2~pixel pitch; (c) 3~px pitch. In (d), the modulation transfer function is measured via the dark-to-light contrast of images such as those in (a)-(c). (e) The point spread function of the optical system, as obtained by the Fourier transform of (d). (f) Ray-trace calculation for the full imaging setup. (g) Close-up of ray-trace at the origin. (h) Close-up of the ray-trace at the focal point of the in-vacuum asphere.}
\label{fig:resolution}
\end{figure}

We directly image the face of the SLM with the lens configuration shown in Fig.~\ref{fig:optical_setup}. The SLM is illuminated with up to 5~W of 532~nm light, sourced from a Spectra-Physics Millenia laser, which is expanded to a diameter of 32~mm to fully illuminate the SLM screen. The SLM spatially modulates the polarization, which is converted into a spatial intensity pattern by the polarizing beamsplitter. This pattern is first imaged by a matched achromat set, creating an intermediate image with a magnification of 0.15$\times$. The light is then imaged onto the plane of the atoms using a 250~mm achromat lens and the 0.42~NA asphere, providing a total magnification of 0.036$\times$. We conducted ray-tracing calculations with this set-up, shown in Fig \ref{fig:resolution}(e)-(g), and find we can geometrically resolve a 1~$\mu{}$m feature separation. We observe some curvature of the image plane, as indicated in Fig~\ref{fig:resolution}(h). With this magnification, we obtain a maximum potential depth for $^{87}$Rb of 2.5~$\mu$K.

The resolution of the SLM imaging system is measured outside of the vacuum system prior to installation. A vacuum chamber window was also present in the test setup to mimic the working conditions of the system as close as possible. We image a set of gratings ranging from 10~pixel pitch to 1~pixel pitch (shown in Fig.~\ref{fig:resolution}(a)-(c)) to directly measure the modulation transfer function (MTF), as shown in Fig.~\ref{fig:resolution}(d). The point spread function (PSF) is the Fourier transform of the modulation transfer function~\cite{Hecht2017}. The point spread function, calculated via the zero-padded Fourier transform of interpolated data from Fig.~\ref{fig:resolution}(d), is plotted in the inset of Fig.~\ref{fig:resolution}(d). The first minimum of the Fourier transform occurs at 0.9~$\mu$m, and the full-width at half-maximum of the point spread function is roughly equal at 0.9~$\mu$m. We may define the resolution as the first minimum of the PSF, and we thus obtain a measured resolution of 0.9~$\mu$m. This resolution value agrees well with a different test, in which the resolution of a gap between two 2$\times$2 pixel squares was measured. We found that a 2 pixel gap (corresponding to 1.4~$\mu$m) was resolved well, while a 1 pixel gap (corresponding to 0.7~$\mu$m) was poorly resolved. We therefore state the resolution as ${0.9\pm 0.2}$~$\mu$m.

\section{2D Degenerate Quantum Gas Production}

\begin{figure*}[ht]
\centering
\includegraphics[width = 450px,keepaspectratio]{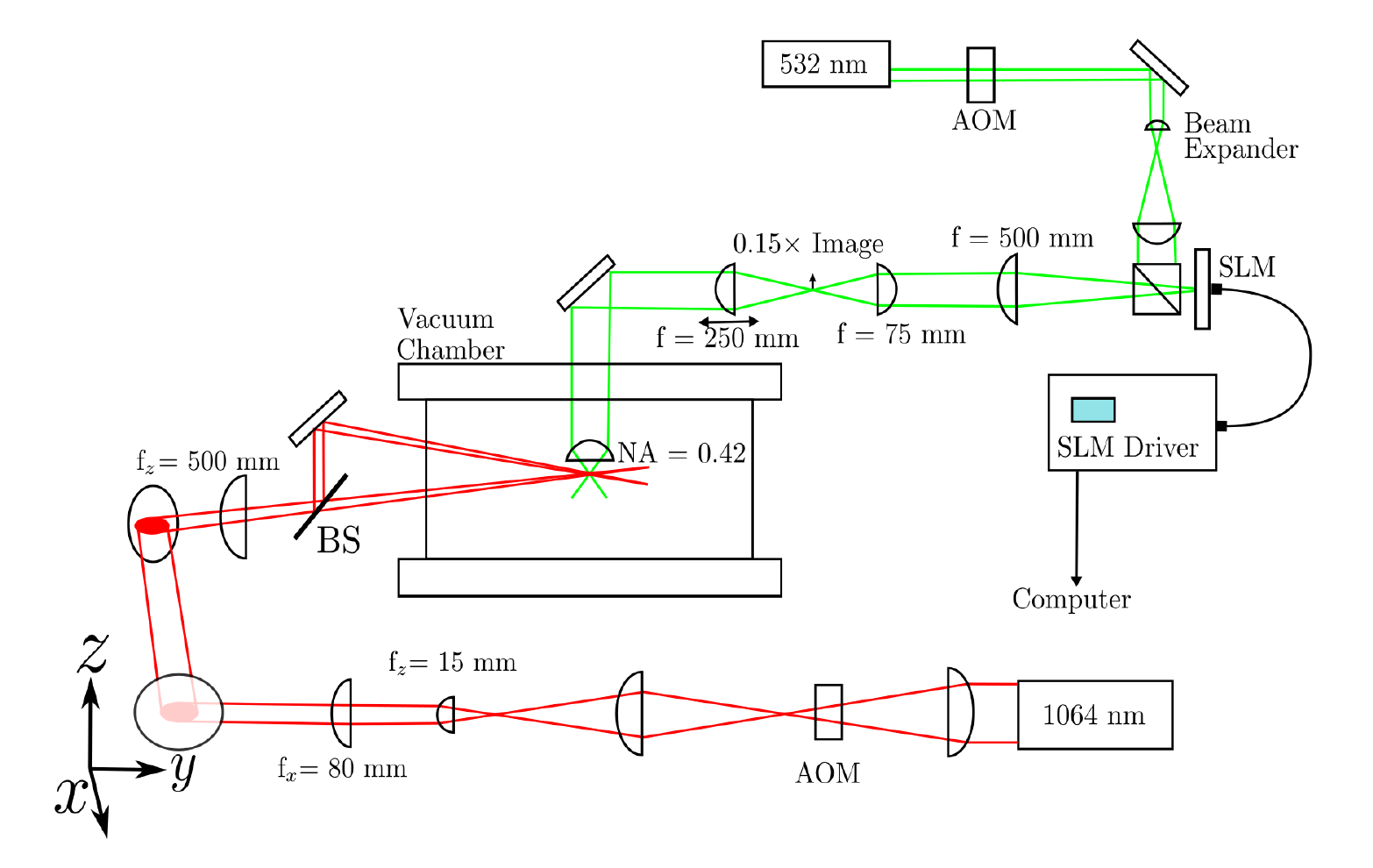}
\caption{
Overview of the optical setup showing the focal lengths and positions of lenses used. The 532~nm light that is used to illuminate the SLM is shown in green. Positioning of the lens inside the vacuum chamber is controlled by bellows (not drawn). The f = 250~mm lens before the chamber is placed on a translation stage for fine control of focusing. An intermediate image between the f = 500~mm lens and the f = 250~mm lens is created with 0.15$\times$ magnification. The 1064~nm light used to produce the 2D trap is shown in red. The beam, emitted by a fiber laser, is power controlled using an AOM . The beam is expanded and then shaped using two collimating cylindrical lenses, the first of 15~mm in focal length in $z$ and then a 80~mm focal length in $x$. The resultant beam is collimated and highly elliptical with an aspect ratio of 16:3 ($x$:$z$). The beam is then angled 3$^{\circ}$ upwards and passes through a weak cylindrical lens with focal length of 500~mm in $z$. A plate beamsplitter (BS) splits the beam creating a vertical beam that is then reflected to intersect the lower beam. Both beams focus and intersect on the atoms creating the 2D trap.}
\label{fig:optical_setup}
\end{figure*}

\begin{figure}[ht]
\includegraphics[width = 3.3in, keepaspectratio]{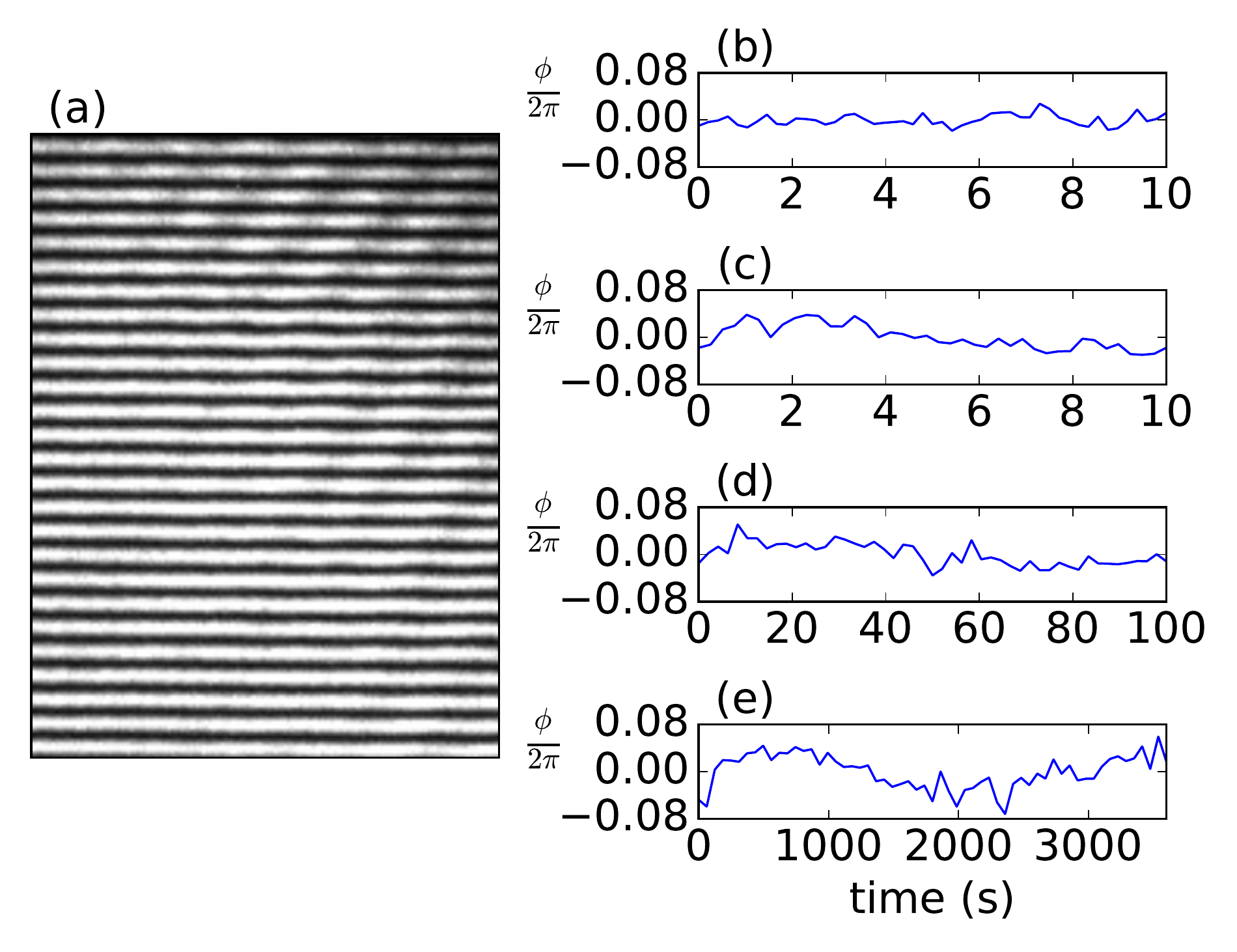}
\caption{(a) Image of the fringes generated by the interfering 1064 nm elliptical beams. The fringes are monitored over time and fitted with a cosine fit where we plot the change in phase taking an image every (b) 200~ms for 10~s (c) 200~ms for 10~s with a source of acoustic noise, namely the finale of Tchaikovsky's 1812 Overture played at 80~dB (d) 2~s for 100~s and (e) 60~s for 3600~s.}
\label{fig:fringe_stability}
\end{figure}

The 2D trap is created by two highly elliptical beams with a wavelength of 1064~nm and an aspect ratio of 16:3 incident at relative angle of 6$^{\circ}$, as shown in Fig.~\ref{fig:optical_setup}. The 1064~nm light is sourced from a Nufern 15~W fiber amplifier (NUA-1064-PB-0015-C0), seeded by an Eagleyard 1064~nm diode laser (EYP-RWS-1064-00080-1500-SOT02-0000). The resulting interference pattern creates sheets of light with a spacing of 8~$\mu{}$m. Each of these light sheets provide weak trapping in the $x$ and $y$ dimensions with trap frequencies of 1~Hz. At the same time, the trap provides a strong confinement in the vertical $z$ dimension with a frequency of 810 $\pm$ 15~Hz. The interference fringes are stable over long periods of time. Figure~\ref{fig:fringe_stability} shows the fringe shift to be less than 15$^{\circ}$ over 1~hour, with minimal high frequency noise and minimal associated parametric heating losses. Each fringe has a large spatial extent with a Rayleigh range of 1.6~mm in the horizontal dimensions. The main fringe has a trap depth of 2~$\mu$K~$\times$~$k_b$. This spatial range enables our experiment to observe transport effects in systems that are characterized by spatial lengths on the order of 200~$\mu{}$m or larger. The ratio $\hbar\omega_z/k_bT = 4.8$ shows that vertical excitations are minimal and atoms trapped in the plane are in a 2D regime. 

The 2D degenerate gas of ultracold atoms is generated by loading an initial Bose-Einstein condensate (BEC) into the 2D trap. Our vacuum chamber setup and BEC generation is described in earlier work~\cite{Wenas}, which we have modified by introducing a 2D MOT for the atom loading stage. In brief, a 2D MOT loads the 3D MOT located inside the main ultra-high vacuum chamber. An all-optical $^{87}$Rb BEC is produced by evaporatively cooling atoms, optically pumped to the $|F=1, m_F =-1 \rangle{}$ state, loaded into a crossed-beam CO$_2$ laser dipole trap. 

Loading of the 2D planar trap begins following the production of the BEC. The ramp sequence is shown in Fig.~\ref{fig:beams_and_ramp}. In the first section of the ramp, the 1064~nm beams are linearly ramped up to full power in 1000~ms, while the CO$_2$ beams are simultaneously ramped down to 20\% of the power at Bose-Einstein condensation. All laser powers are maintained at this power for 100 ms to equilibrate the system, with the CO$_2$ beams providing weak radial confinement and the 1064 nm beams providing suspension against gravity. The CO$_2$ beams are subsequently ramped down to zero in 20~ms, which releases the atoms into the 2D trap. Using this loading scheme we obtain a loading efficiency of 70~$\pm$~1\% into the 2D trap. A maximum of two planar traps are loaded at any time, because the initial 1~$\mu$m size of the BEC is significantly smaller than planar spacing of 8~$\mu$m. Gravity assists in ensuring that a single planar trap is always predominantly loaded.

\begin{figure}[ht]
\centering
\includegraphics[width = 240px,keepaspectratio]{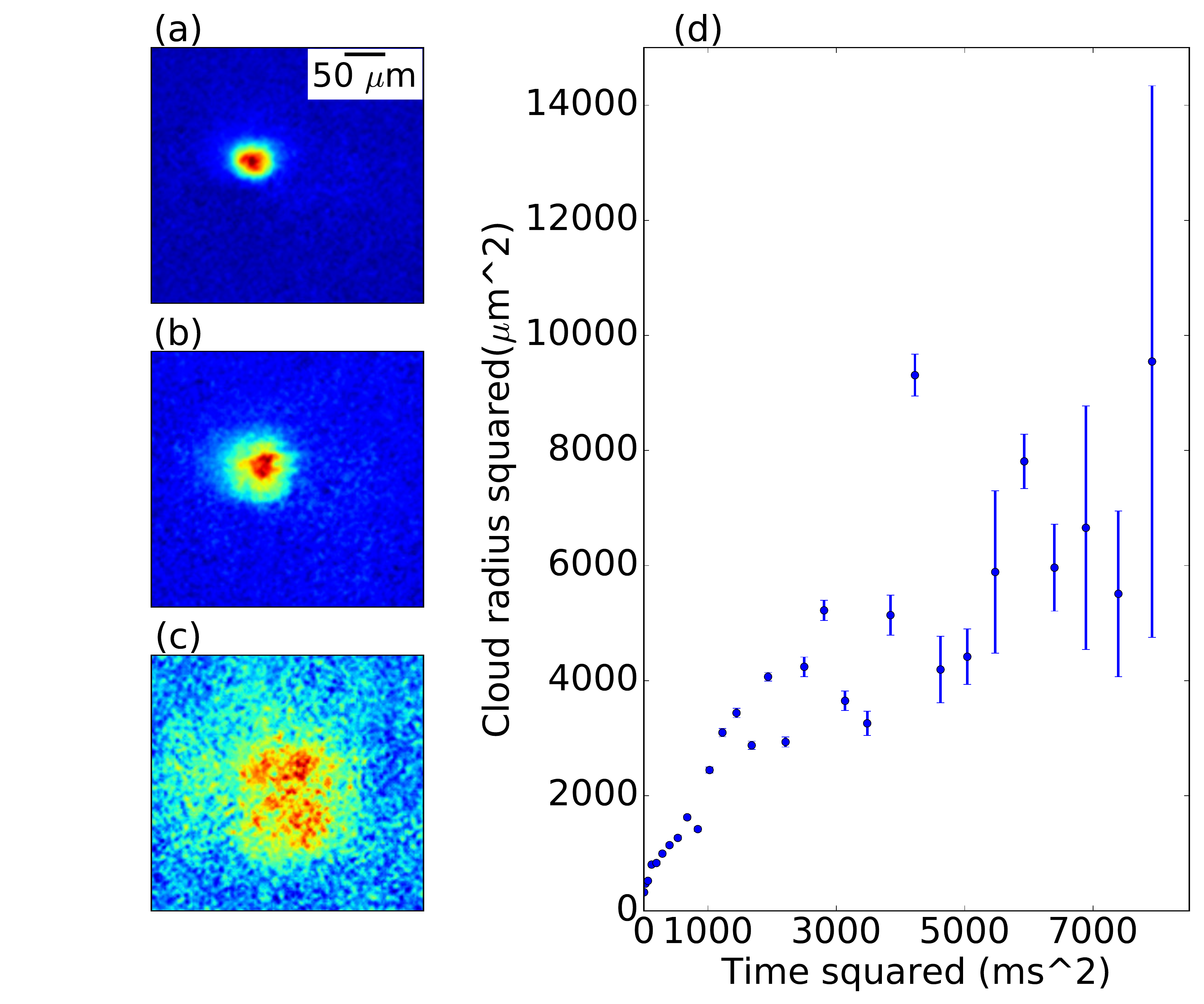}
\caption{Expansion of the atoms in the 2D trap. Images of the atoms taken after (a) 8~ms, (b) 20~ms and (c) 50~ms expansion times averaged over 3 experimental runs.The atomic cloud profiles are fitted with Gaussian functions from which the atomic radius is extracted. The atomic radius is plotted with time (d) for 100~ms of total expansion. The linear increase of the radial size shows ballistic expansion of the cloud in the 2D trap.}
\label{fig:atom_expansion}
\end{figure}

We now characterise the behavior of the 2D degenerate gas of ultracold atoms inside the 2D trap. After the atoms are loaded into the 2D trap they are left to expand. Fig.~\ref{fig:atom_expansion}(a--c) shows absorption images of atoms expanding inside the 2D trap for different expansion times. For ballistic expansion, we expect the cloud radius to obey $R^2(t) = R^2(0) + (k_BT/m)t^2$, in which $t$ is time, $T$ is the temperature of the atom cloud and $m$ is the mass of a $^{87}$Rb atom. We measure the atomic radius at different expansion times and observe a linear relationship between $R^2$ and $t^2$ over a time period greater than 100~ms, with the plot shown in Fig.~\ref{fig:atom_expansion}(d). By applying a linear fit to the data we obtain a temperature of $8\pm 0.5$~nK for the atoms corresponding to a de Broglie wavelength of 2~$\mu{}$m. A measurement of atom retention by the 2D trap gives a trap lifetime of measurement 2~s. It should be stressed that any heating due to vibrations or otherwise will cause the atoms to escape from the trap, which has a shallow trap-depth under gravity of 2~$\mu$K.

\section{Imaging}

\begin{figure}[ht]
\centering
\includegraphics[width=240px,keepaspectratio]{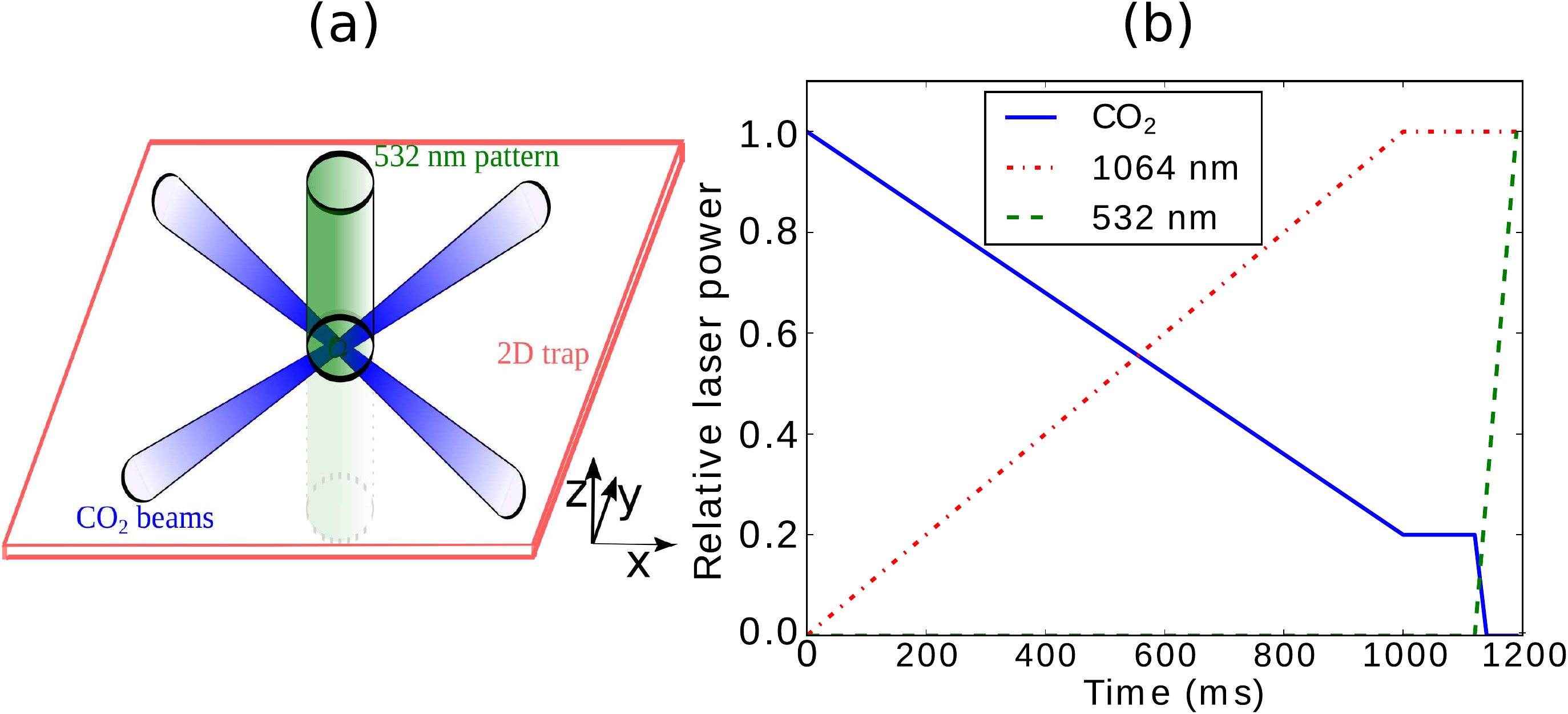}
\caption{(a) Representation of the intersection of all 3 beams; the CO$_2$ beams (blue); 1064~nm planar trap (red); and 532~nm SLM generated pattern (green). In this case the pattern imaged is a hollow cylinder. (b) Beam intensity ramps for all 3 beams during the loading process and application of the tailored potential. A relative laser power of 1 corresponds to an output of 50~mW for the CO$_2$ laser, 8~W for the 1064~nm laser and 3~W for the 532~nm laser.}
\label{fig:beams_and_ramp}
\end{figure}

During the CO$_2$ laser ramp-down, the 532~nm light is ramped to full power simultaneously. The 532~nm light projects the desired potential onto the atoms. In this section we show some example experiments which reflect the control over atoms which the setup provides. In these experiments, the atoms begin from a point source at the location of the CO$_2$ beam intersection. A visualization of the location of all 3 beams and intensity ramps is shown in Fig.~\ref{fig:beams_and_ramp}. With blue-detuned light, a clear method is to image the outline of a certain shape and allow the atoms to expand to fill the void. Some previous experiments have found it useful to evaporate the atoms in the presence of the potential, such that the ground state of the projected potential becomes occupied~\cite{Gauthier}. This experiment differs slightly, in that the atoms begin expansion from a point source within the projected potential. This is a transmissive experiment, and an open path to the origin is required for atom transport to occur to a given point in space.

\begin{figure}[ht]
\centering
\includegraphics[width = 200 px, keepaspectratio]{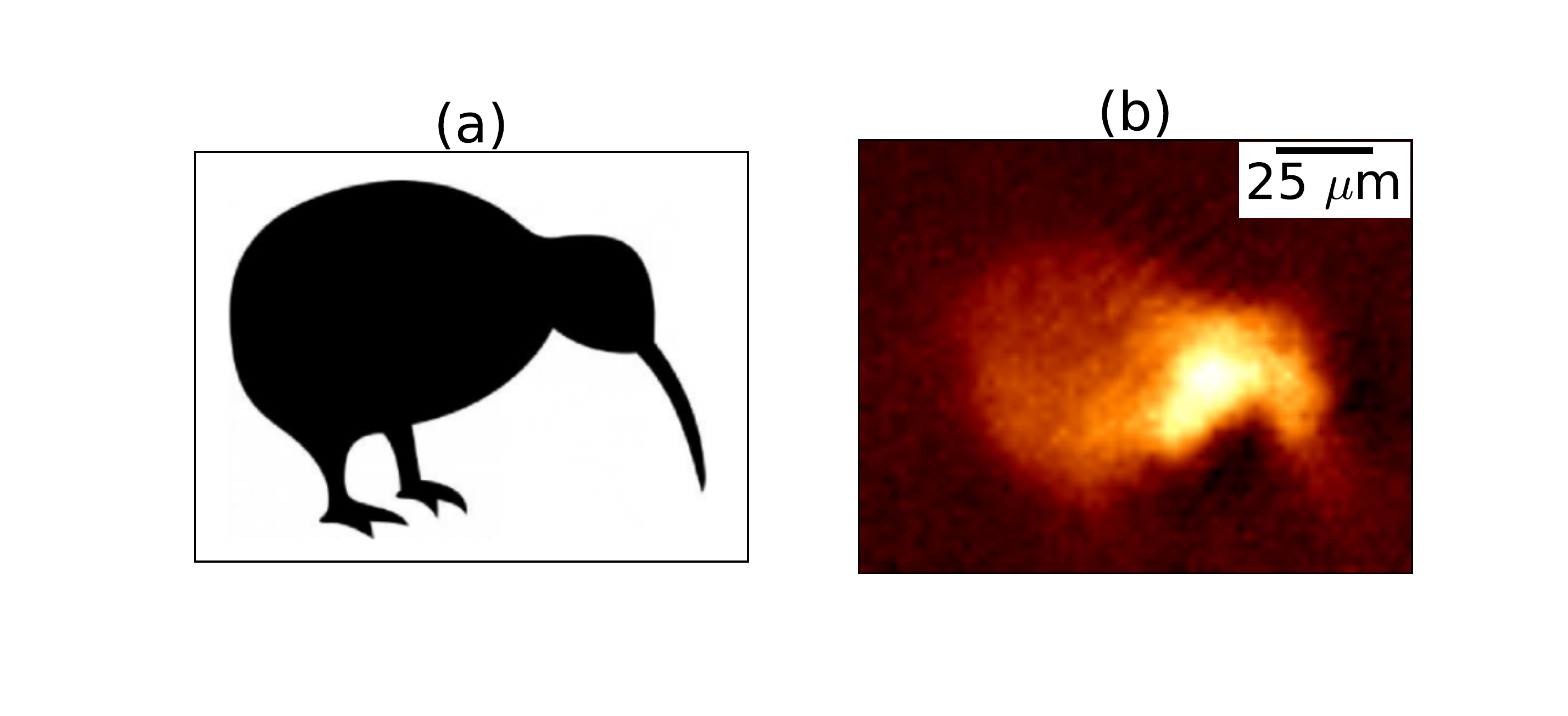}
\caption{Atoms taking on the shape of a kiwi-bird potential, which is 150~$\mu$m head to tail. (a) Image of the kiwi shown on the SLM (b) Image of atoms inside the kiwi potential after 300 ms of expansion averaged over 20 experimental runs.}
\label{fig:kiwi}
\end{figure}

Figure~\ref{fig:kiwi} shows an image of atoms where the outline of a kiwi bird has been imaged onto the plane. The atoms are imaged by a second 4-f imaging system of NA 0.23. The general shape outline of the kiwi is well reproduced by the atomic density profile. The atom propagation into the smaller `beak' and `feet' structures is limited by mode-matching requirements, which are strict given that the de Broglie wavelength of approximately 2~$\mu$m is comparable to the channel width of 5~$\mu$m.

\begin{figure}[ht]
\centering
\includegraphics[width = 250px, keepaspectratio]{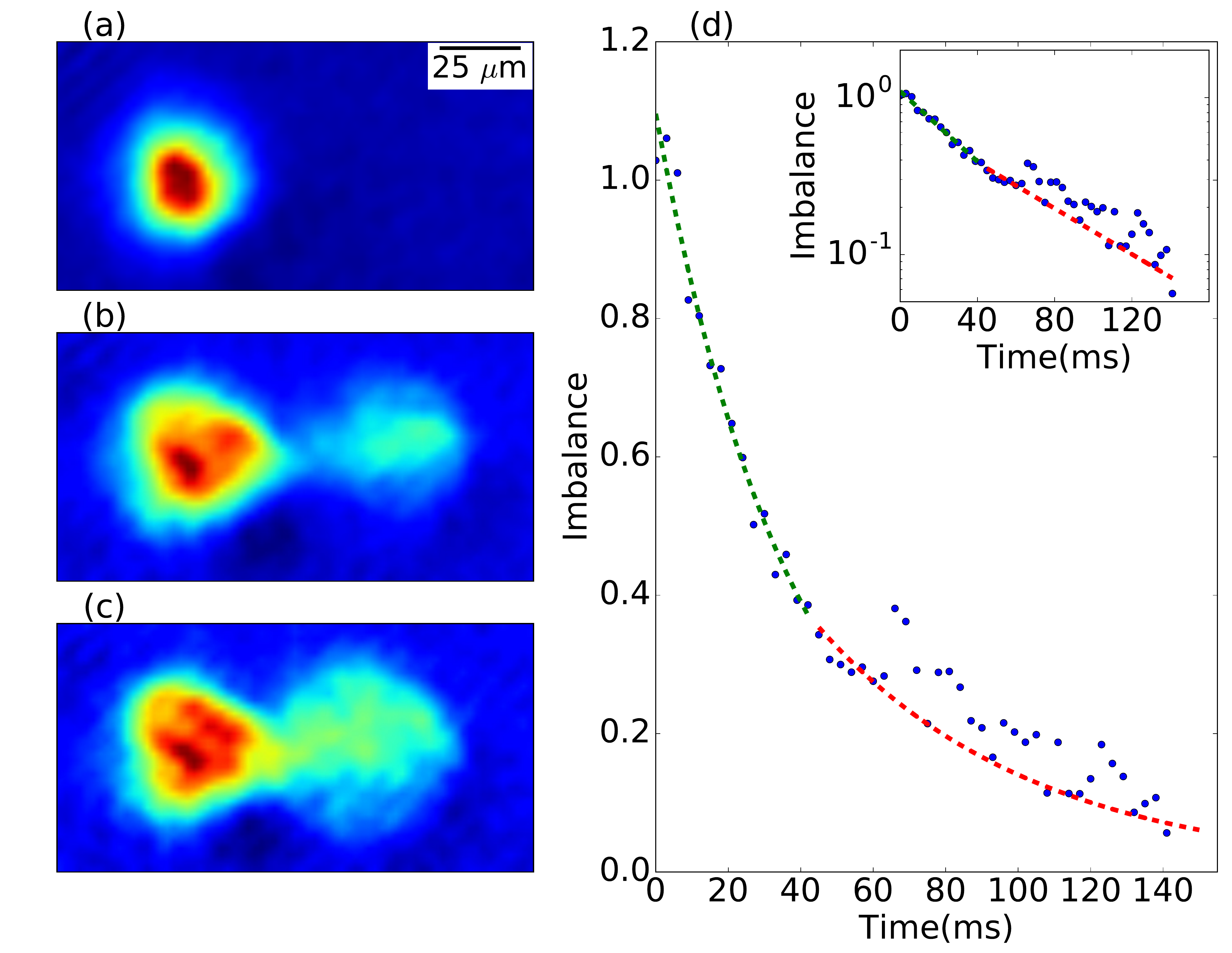}
\caption{(a) Initial loading of atoms into a single reservoir on the left-hand-side of the `dumbbell' shape. The channel length is 35~$\mu$m and the width is 30~$\mu$m. The reservoir radii are 35~$\mu$m.  (b) Expansion of the atoms inside the trap after 45~ms. (c) Atoms loading into the secondary reservoir of the dumbbell after 120~ms. (d) Plot of number imbalance between both reservoirs in time. The two transport regimes, (green) ballistic and (red) slow decay coupled to oscillations, are fitted. Inset: Semi-logarithmic plot of the same data, showing the two regimes.}
\label{fig:dumbell}
\end{figure}

We can also look at more functional geometries to analyze atomic transport dynamics. A `dumbbell' structure, where an initial circular reservoir loaded with the atoms is connected by a channel to a secondary reservoir, allows for a controlled study of transmissive experiments for where the chemical potential drives the atoms from one reservoir to the other. Figure \ref{fig:dumbell} shows the expansion of the atoms into a such a potential with a 35 $\mu{}$m long channel after 150 ms of expansion time. We define the number imbalance as:

\begin{equation}
\Delta{}N = \frac{N_i - N_f}{N_i+N_f},
\end{equation} 

\noindent{}where $N_i$ is the atom number in the initial reservoir and $N_f$ of the final reservoir. Dumbbell structures create a double-well structure which can be used to model a range of systems. For example, dumbbell geometries have been used to model a resistor-capacitor-inductor (RCL) atomtronic circuit~\cite{Eckle,PhysRevA.94.023626,ElectricAnalogsAtoms}. We create an outline of a dumbbell using the SLM, with the SLM versatility allowing for the channel width and length to be tailored to investigate the atom transport. The data in Fig.~\ref{fig:dumbell}(d) shows the atom number imbalance against time. Two transport regimes are observed. The first 40~ms are dominated by ballistic transport, resulting in a rapid linear decrease of the imbalance. Once a sufficient number of atoms have entered the second reservoir, the effects of the chemical potential dominate the transport. We observe an exponential decay coupled with oscillatory behaviour. In this regime the system is best modeled by the above RCL atomtronic circuit. This model predicts oscillatory behaviour and energy exchange between the chemical potential of the reservoirs and the kinetic energy of the atoms, at a frequency determined by $\omega=1/\sqrt{LC}$. 

In future experiments, adding disorder to the channel opens up the possibility of studying localization effects.  The long spatial range of our experiment will also allow us to study length scales comparable to that of localization lengths for cold atom systems. 

\begin{figure}[ht]
\centering
\includegraphics[width=200px,keepaspectratio]{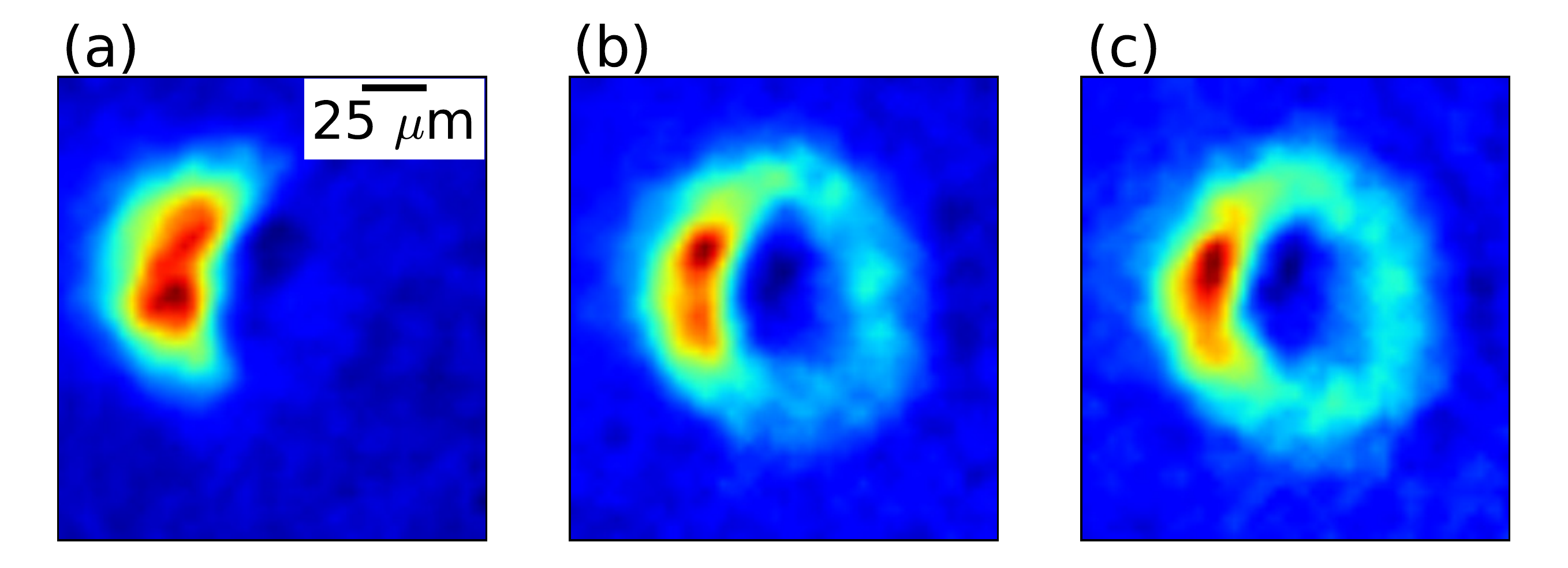}
\caption{Atoms expanding into a ring potential. The inner diameter of the ring is 43~$\mu$m (60 SLM pixels) and the outer diameter is 76~$\mu$m (120 SLM pixels). The atoms are shown at different expansion times of (a) 0~ms (b) 50~ms (c) 100~ms. }
\label{fig:rings}
\end{figure}

Other geometries can be used, such as ring traps, as shown in Fig.~\ref{fig:rings}. Here atoms are initially loaded onto the left-side of a 70~$\mu$m outer diameter ring and allowed to expand. Ring traps have proven to be a test bed for numerous quantum effects \cite{ring_trap_Cassettari_group}. Moreover, there is the possibility to use rings as atomtronic systems by analyzing the atom number imbalances in the left and right side of the ring~\cite{RingHysterisis}.   

\section{Summary}

To summarize, we have presented an ultracold atom setup capable of conducting transport experiments over large length ranges in 2D. Customizable repulsive potentials are generated by a SLM using intensity modulation to shape the light field of a 532~nm laser. An imaging system with an in-vacuo lens is used to generate optical potentials with a resolution of $0.9\pm 0.2$~$\mu$m. 
Our initial results with a variety of geometries demonstrate the versatility of our setup. Our system provides a test bed for experiments on quantum transport and localization, as well as atomtronics.
	
\section{Acknowledgments}
We gratefully acknowledge the support of the Marsden Fund, with funding
administered by the Royal Society of New Zealand.


\begin{thebibliography}{43}%
	\makeatletter
	\providecommand \@ifxundefined [1]{%
		\@ifx{#1\undefined}
	}%
	\providecommand \@ifnum [1]{%
		\ifnum #1\expandafter \@firstoftwo
		\else \expandafter \@secondoftwo
		\fi
	}%
	\providecommand \@ifx [1]{%
		\ifx #1\expandafter \@firstoftwo
		\else \expandafter \@secondoftwo
		\fi
	}%
	\providecommand \natexlab [1]{#1}%
	\providecommand \enquote  [1]{``#1''}%
	\providecommand \bibnamefont  [1]{#1}%
	\providecommand \bibfnamefont [1]{#1}%
	\providecommand \citenamefont [1]{#1}%
	\providecommand \href@noop [0]{\@secondoftwo}%
	\providecommand \href [0]{\begingroup \@sanitize@url \@href}%
	\providecommand \@href[1]{\@@startlink{#1}\@@href}%
	\providecommand \@@href[1]{\endgroup#1\@@endlink}%
	\providecommand \@sanitize@url [0]{\catcode `\\12\catcode `\$12\catcode
		`\&12\catcode `\#12\catcode `\^12\catcode `\_12\catcode `\%12\relax}%
	\providecommand \@@startlink[1]{}%
	\providecommand \@@endlink[0]{}%
	\providecommand \url  [0]{\begingroup\@sanitize@url \@url }%
	\providecommand \@url [1]{\endgroup\@href {#1}{\urlprefix }}%
	\providecommand \urlprefix  [0]{URL }%
	\providecommand \Eprint [0]{\href }%
	\providecommand \doibase [0]{http://dx.doi.org/}%
	\providecommand \selectlanguage [0]{\@gobble}%
	\providecommand \bibinfo  [0]{\@secondoftwo}%
	\providecommand \bibfield  [0]{\@secondoftwo}%
	\providecommand \translation [1]{[#1]}%
	\providecommand \BibitemOpen [0]{}%
	\providecommand \bibitemStop [0]{}%
	\providecommand \bibitemNoStop [0]{.\EOS\space}%
	\providecommand \EOS [0]{\spacefactor3000\relax}%
	\providecommand \BibitemShut  [1]{\csname bibitem#1\endcsname}%
	\let\auto@bib@innerbib\@empty
	\bibitem [{\citenamefont {Bloch}, \citenamefont {Dalibard},\ and\ \citenamefont
		{Nascimb\`{e}ne}(2012)}]{Bloch2012}%
	\BibitemOpen
	\bibfield  {author} {\bibinfo {author} {\bibfnamefont {I.}~\bibnamefont
			{Bloch}}, \bibinfo {author} {\bibfnamefont {J.}~\bibnamefont {Dalibard}}, \
		and\ \bibinfo {author} {\bibfnamefont {S.}~\bibnamefont {Nascimb\`{e}ne}},\
	}\href {http://dx.doi.org/10.1038/nphys2259} {\bibfield  {journal} {\bibinfo
		{journal} {Nat. Phys.}\ }\textbf {\bibinfo {volume} {8}},\ \bibinfo {pages}
	{267} (\bibinfo {year} {2012})}\BibitemShut {NoStop}%
\bibitem [{\citenamefont {O'Malley}\ \emph {et~al.}(2016)\citenamefont
	{O'Malley} \emph {et~al.}}]{Omalley2016}%
\BibitemOpen
\bibfield  {author} {\bibinfo {author} {\bibfnamefont {P.~J.~J.}\
		\bibnamefont {O'Malley}} \emph {et~al.},\ }\href
{http://link.aps.org/doi/10.1103/PhysRevX.6.031007} {\bibfield  {journal}
	{\bibinfo  {journal} {Phys. Rev. X}\ }\textbf {\bibinfo {volume} {6}},\
	\bibinfo {pages} {031007} (\bibinfo {year} {2016})}\BibitemShut {NoStop}%
\bibitem [{\citenamefont {Cirac}\ and\ \citenamefont
	{Zoller}(2012)}]{Cirac2012}%
\BibitemOpen
\bibfield  {author} {\bibinfo {author} {\bibfnamefont {J.~I.}\ \bibnamefont
		{Cirac}}\ and\ \bibinfo {author} {\bibfnamefont {P.}~\bibnamefont {Zoller}},\
}\href {\doibase 10.1038/nphys2275} {\bibfield  {journal} {\bibinfo
	{journal} {Nat. Phys.}\ }\textbf {\bibinfo {volume} {8}},\ \bibinfo {pages}
{264} (\bibinfo {year} {2012})}\BibitemShut {NoStop}%
\bibitem [{\citenamefont {Lanyon}\ \emph {et~al.}(2011)\citenamefont {Lanyon}
	\emph {et~al.}}]{Lanyon2011}%
\BibitemOpen
\bibfield  {author} {\bibinfo {author} {\bibfnamefont {B.~P.}\ \bibnamefont
		{Lanyon}} \emph {et~al.},\ }\href {\doibase 10.1126/science.1208001}
{\bibfield  {journal} {\bibinfo  {journal} {Science}\ }\textbf {\bibinfo
		{volume} {334}},\ \bibinfo {pages} {57} (\bibinfo {year} {2011})}\BibitemShut
{NoStop}%
\bibitem [{\citenamefont {Sherson}\ \emph {et~al.}(2010)\citenamefont
	{Sherson}, \citenamefont {Weitenberg}, \citenamefont {Endres}, \citenamefont
	{Cheneau}, \citenamefont {Bloch},\ and\ \citenamefont {Kuhr}}]{Sherson2010}%
\BibitemOpen
\bibfield  {author} {\bibinfo {author} {\bibfnamefont {J.~F.}\ \bibnamefont
		{Sherson}}, \bibinfo {author} {\bibfnamefont {C.}~\bibnamefont {Weitenberg}},
	\bibinfo {author} {\bibfnamefont {M.}~\bibnamefont {Endres}}, \bibinfo
	{author} {\bibfnamefont {M.}~\bibnamefont {Cheneau}}, \bibinfo {author}
	{\bibfnamefont {I.}~\bibnamefont {Bloch}}, \ and\ \bibinfo {author}
	{\bibfnamefont {S.}~\bibnamefont {Kuhr}},\ }\href {\doibase
	10.1038/nature09378} {\bibfield  {journal} {\bibinfo  {journal} {Nature}\
	}\textbf {\bibinfo {volume} {467}},\ \bibinfo {pages} {68} (\bibinfo {year}
	{2010})}\BibitemShut {NoStop}%
\bibitem [{\citenamefont {Chien}, \citenamefont {Peotta},\ and\ \citenamefont
	{Di~Ventra}(2015)}]{Chien2015}%
\BibitemOpen
\bibfield  {author} {\bibinfo {author} {\bibfnamefont {C.-C.}\ \bibnamefont
		{Chien}}, \bibinfo {author} {\bibfnamefont {S.}~\bibnamefont {Peotta}}, \
	and\ \bibinfo {author} {\bibfnamefont {M.}~\bibnamefont {Di~Ventra}},\ }\href
{http://dx.doi.org/10.1038/nphys3531} {\bibfield  {journal} {\bibinfo
		{journal} {Nat. Phys.}\ }\textbf {\bibinfo {volume} {11}},\ \bibinfo {pages}
	{998} (\bibinfo {year} {2015})}\BibitemShut {NoStop}%
\bibitem [{\citenamefont {Billy}\ \emph {et~al.}(2008)\citenamefont {Billy},
	\citenamefont {Josse}, \citenamefont {Zuo}, \citenamefont {Bernard},
	\citenamefont {Hambrecht}, \citenamefont {Lugan}, \citenamefont
	{Cl\'{e}ment}, \citenamefont {Sanchez-Palencia}, \citenamefont {Bouyer},\
	and\ \citenamefont {Aspect}}]{Billy2008}%
\BibitemOpen
\bibfield  {author} {\bibinfo {author} {\bibfnamefont {J.}~\bibnamefont
		{Billy}}, \bibinfo {author} {\bibfnamefont {V.}~\bibnamefont {Josse}},
	\bibinfo {author} {\bibfnamefont {Z.}~\bibnamefont {Zuo}}, \bibinfo {author}
	{\bibfnamefont {A.}~\bibnamefont {Bernard}}, \bibinfo {author} {\bibfnamefont
		{B.}~\bibnamefont {Hambrecht}}, \bibinfo {author} {\bibfnamefont
		{P.}~\bibnamefont {Lugan}}, \bibinfo {author} {\bibfnamefont
		{D.}~\bibnamefont {Cl\'{e}ment}}, \bibinfo {author} {\bibfnamefont
		{L.}~\bibnamefont {Sanchez-Palencia}}, \bibinfo {author} {\bibfnamefont
		{P.}~\bibnamefont {Bouyer}}, \ and\ \bibinfo {author} {\bibfnamefont
		{A.}~\bibnamefont {Aspect}},\ }\href {http://dx.doi.org/10.1038/nature07000}
{\bibfield  {journal} {\bibinfo  {journal} {Nature}\ }\textbf {\bibinfo
		{volume} {453}},\ \bibinfo {pages} {891} (\bibinfo {year}
	{2008})}\BibitemShut {NoStop}%
\bibitem [{\citenamefont {Zupancic}\ \emph {et~al.}(2016)\citenamefont
	{Zupancic}, \citenamefont {Preiss}, \citenamefont {Ma}, \citenamefont
	{Lukin}, \citenamefont {Tai}, \citenamefont {Rispoli}, \citenamefont
	{Islam},\ and\ \citenamefont {Greiner}}]{Greiner_DMD_phase}%
\BibitemOpen
\bibfield  {author} {\bibinfo {author} {\bibfnamefont {P.}~\bibnamefont
		{Zupancic}}, \bibinfo {author} {\bibfnamefont {P.~M.}\ \bibnamefont
		{Preiss}}, \bibinfo {author} {\bibfnamefont {R.}~\bibnamefont {Ma}}, \bibinfo
	{author} {\bibfnamefont {A.}~\bibnamefont {Lukin}}, \bibinfo {author}
	{\bibfnamefont {M.~E.}\ \bibnamefont {Tai}}, \bibinfo {author} {\bibfnamefont
		{M.}~\bibnamefont {Rispoli}}, \bibinfo {author} {\bibfnamefont
		{R.}~\bibnamefont {Islam}}, \ and\ \bibinfo {author} {\bibfnamefont
		{M.}~\bibnamefont {Greiner}},\ }\href {\doibase 10.1364/OE.24.013881}
{\bibfield  {journal} {\bibinfo  {journal} {Opt. Express}\ }\textbf {\bibinfo
		{volume} {24}},\ \bibinfo {pages} {13881} (\bibinfo {year}
	{2016})}\BibitemShut {NoStop}%
\bibitem [{\citenamefont {Gauthier}\ \emph {et~al.}(2016)\citenamefont
	{Gauthier}, \citenamefont {Lenton}, \citenamefont {Parry}, \citenamefont
	{Baker}, \citenamefont {Davis}, \citenamefont {Rubinsztein-Dunlop},\ and\
	\citenamefont {Neely}}]{Gauthier}%
\BibitemOpen
\bibfield  {author} {\bibinfo {author} {\bibfnamefont {G.}~\bibnamefont
		{Gauthier}}, \bibinfo {author} {\bibfnamefont {I.}~\bibnamefont {Lenton}},
	\bibinfo {author} {\bibfnamefont {N.~M.}\ \bibnamefont {Parry}}, \bibinfo
	{author} {\bibfnamefont {M.}~\bibnamefont {Baker}}, \bibinfo {author}
	{\bibfnamefont {M.~J.}\ \bibnamefont {Davis}}, \bibinfo {author}
	{\bibfnamefont {H.}~\bibnamefont {Rubinsztein-Dunlop}}, \ and\ \bibinfo
	{author} {\bibfnamefont {T.~W.}\ \bibnamefont {Neely}},\ }\href {\doibase
	10.1364/OPTICA.3.001136} {\bibfield  {journal} {\bibinfo  {journal} {Optica}\
	}\textbf {\bibinfo {volume} {3}},\ \bibinfo {pages} {1136} (\bibinfo {year}
	{2016})}\BibitemShut {NoStop}%
\bibitem [{\citenamefont {Muldoon}\ \emph {et~al.}(2012)\citenamefont
	{Muldoon}, \citenamefont {Brandt}, \citenamefont {Dong}, \citenamefont
	{Stuart}, \citenamefont {Brainis}, \citenamefont {Himsworth},\ and\
	\citenamefont {Kuhn}}]{optical_tweezers_atoms}%
\BibitemOpen
\bibfield  {author} {\bibinfo {author} {\bibfnamefont {C.}~\bibnamefont
		{Muldoon}}, \bibinfo {author} {\bibfnamefont {L.}~\bibnamefont {Brandt}},
	\bibinfo {author} {\bibfnamefont {J.}~\bibnamefont {Dong}}, \bibinfo {author}
	{\bibfnamefont {D.}~\bibnamefont {Stuart}}, \bibinfo {author} {\bibfnamefont
		{E.}~\bibnamefont {Brainis}}, \bibinfo {author} {\bibfnamefont
		{M.}~\bibnamefont {Himsworth}}, \ and\ \bibinfo {author} {\bibfnamefont
		{A.}~\bibnamefont {Kuhn}},\ }\href@noop {} {\bibfield  {journal} {\bibinfo
		{journal} {New Journal of Physics}\ }\textbf {\bibinfo {volume} {14}},\
	\bibinfo {pages} {073051} (\bibinfo {year} {2012})}\BibitemShut {NoStop}%
\bibitem [{\citenamefont {Liang}\ \emph {et~al.}(2009)\citenamefont {Liang},
	\citenamefont {Rudolph N.~Kohn}, \citenamefont {Becker},\ and\ \citenamefont
	{Heinzen}}]{Liang:09}%
\BibitemOpen
\bibfield  {author} {\bibinfo {author} {\bibfnamefont {J.}~\bibnamefont
		{Liang}}, \bibinfo {author} {\bibfnamefont {J.}~\bibnamefont {Rudolph
			N.~Kohn}}, \bibinfo {author} {\bibfnamefont {M.~F.}\ \bibnamefont {Becker}},
	\ and\ \bibinfo {author} {\bibfnamefont {D.~J.}\ \bibnamefont {Heinzen}},\
}\href {\doibase 10.1364/AO.48.001955} {\bibfield  {journal} {\bibinfo
	{journal} {Appl. Opt.}\ }\textbf {\bibinfo {volume} {48}},\ \bibinfo {pages}
{1955} (\bibinfo {year} {2009})}\BibitemShut {NoStop}%
\bibitem [{\citenamefont {Henderson}\ \emph {et~al.}(2009)\citenamefont
	{Henderson}, \citenamefont {Ryu}, \citenamefont {MacCormick},\ and\
	\citenamefont {Boshier}}]{painted_potentials}%
\BibitemOpen
\bibfield  {author} {\bibinfo {author} {\bibfnamefont {K.}~\bibnamefont
		{Henderson}}, \bibinfo {author} {\bibfnamefont {C.}~\bibnamefont {Ryu}},
	\bibinfo {author} {\bibfnamefont {C.}~\bibnamefont {MacCormick}}, \ and\
	\bibinfo {author} {\bibfnamefont {M.~G.}\ \bibnamefont {Boshier}},\ }\href
{http://stacks.iop.org/1367-2630/11/i=4/a=043030} {\bibfield  {journal}
	{\bibinfo  {journal} {New Journal of Physics}\ }\textbf {\bibinfo {volume}
		{11}},\ \bibinfo {pages} {043030} (\bibinfo {year} {2009})}\BibitemShut
{NoStop}%
\bibitem [{\citenamefont {Bell}\ \emph {et~al.}(2016)\citenamefont {Bell},
	\citenamefont {Glidden}, \citenamefont {Humbert}, \citenamefont {Bromley},
	\citenamefont {Haine}, \citenamefont {Davis}, \citenamefont {Neely},
	\citenamefont {Baker},\ and\ \citenamefont
	{Rubinsztein-Dunlop}}]{PaintedRing}%
\BibitemOpen
\bibfield  {author} {\bibinfo {author} {\bibfnamefont {T.~A.}\ \bibnamefont
		{Bell}}, \bibinfo {author} {\bibfnamefont {J.~A.~P.}\ \bibnamefont
		{Glidden}}, \bibinfo {author} {\bibfnamefont {L.}~\bibnamefont {Humbert}},
	\bibinfo {author} {\bibfnamefont {M.~W.~J.}\ \bibnamefont {Bromley}},
	\bibinfo {author} {\bibfnamefont {S.~A.}\ \bibnamefont {Haine}}, \bibinfo
	{author} {\bibfnamefont {M.~J.}\ \bibnamefont {Davis}}, \bibinfo {author}
	{\bibfnamefont {T.~W.}\ \bibnamefont {Neely}}, \bibinfo {author}
	{\bibfnamefont {M.~A.}\ \bibnamefont {Baker}}, \ and\ \bibinfo {author}
	{\bibfnamefont {H.}~\bibnamefont {Rubinsztein-Dunlop}},\ }\href
{http://stacks.iop.org/1367-2630/18/i=3/a=035003} {\bibfield  {journal}
	{\bibinfo  {journal} {New Journal of Physics}\ }\textbf {\bibinfo {volume}
		{18}},\ \bibinfo {pages} {035003} (\bibinfo {year} {2016})}\BibitemShut
{NoStop}%
\bibitem [{\citenamefont {Bakr}\ \emph {et~al.}(2009)\citenamefont {Bakr},
	\citenamefont {Gillen}, \citenamefont {Peng}, \citenamefont {Folling},\ and\
	\citenamefont {Greiner}}]{Greiner_Atom_Microscope}%
\BibitemOpen
\bibfield  {author} {\bibinfo {author} {\bibfnamefont {W.~S.}\ \bibnamefont
		{Bakr}}, \bibinfo {author} {\bibfnamefont {J.~I.}\ \bibnamefont {Gillen}},
	\bibinfo {author} {\bibfnamefont {A.}~\bibnamefont {Peng}}, \bibinfo {author}
	{\bibfnamefont {S.}~\bibnamefont {Folling}}, \ and\ \bibinfo {author}
	{\bibfnamefont {M.}~\bibnamefont {Greiner}},\ }\href {\doibase
	10.1038/nature08482} {\bibfield  {journal} {\bibinfo  {journal} {Nature}\
	}\textbf {\bibinfo {volume} {462}},\ \bibinfo {pages} {74} (\bibinfo {year}
	{2009})}\BibitemShut {NoStop}%
\bibitem [{\citenamefont {Ramanathan}\ \emph {et~al.}(2011)\citenamefont
	{Ramanathan}, \citenamefont {Wright}, \citenamefont {Muniz}, \citenamefont
	{Zelan}, \citenamefont {Hill}, \citenamefont {Lobb}, \citenamefont
	{Helmerson}, \citenamefont {Phillips},\ and\ \citenamefont
	{Campbell}}]{ToroidalSuperflow}%
\BibitemOpen
\bibfield  {author} {\bibinfo {author} {\bibfnamefont {A.}~\bibnamefont
		{Ramanathan}}, \bibinfo {author} {\bibfnamefont {K.~C.}\ \bibnamefont
		{Wright}}, \bibinfo {author} {\bibfnamefont {S.~R.}\ \bibnamefont {Muniz}},
	\bibinfo {author} {\bibfnamefont {M.}~\bibnamefont {Zelan}}, \bibinfo
	{author} {\bibfnamefont {W.~T.}\ \bibnamefont {Hill}}, \bibinfo {author}
	{\bibfnamefont {C.~J.}\ \bibnamefont {Lobb}}, \bibinfo {author}
	{\bibfnamefont {K.}~\bibnamefont {Helmerson}}, \bibinfo {author}
	{\bibfnamefont {W.~D.}\ \bibnamefont {Phillips}}, \ and\ \bibinfo {author}
	{\bibfnamefont {G.~K.}\ \bibnamefont {Campbell}},\ }\href {\doibase
	10.1103/PhysRevLett.106.130401} {\bibfield  {journal} {\bibinfo  {journal}
		{Phys. Rev. Lett.}\ }\textbf {\bibinfo {volume} {106}},\ \bibinfo {pages}
	{130401} (\bibinfo {year} {2011})}\BibitemShut {NoStop}%
\bibitem [{\citenamefont {Wright}\ \emph {et~al.}(2013)\citenamefont {Wright},
	\citenamefont {Blakestad}, \citenamefont {Lobb}, \citenamefont {Phillips},\
	and\ \citenamefont {Campbell}}]{StirredRing}%
\BibitemOpen
\bibfield  {author} {\bibinfo {author} {\bibfnamefont {K.~C.}\ \bibnamefont
		{Wright}}, \bibinfo {author} {\bibfnamefont {R.~B.}\ \bibnamefont
		{Blakestad}}, \bibinfo {author} {\bibfnamefont {C.~J.}\ \bibnamefont {Lobb}},
	\bibinfo {author} {\bibfnamefont {W.~D.}\ \bibnamefont {Phillips}}, \ and\
	\bibinfo {author} {\bibfnamefont {G.~K.}\ \bibnamefont {Campbell}},\ }\href
{\doibase 10.1103/PhysRevA.88.063633} {\bibfield  {journal} {\bibinfo
		{journal} {Phys. Rev. A}\ }\textbf {\bibinfo {volume} {88}},\ \bibinfo
	{pages} {063633} (\bibinfo {year} {2013})}\BibitemShut {NoStop}%
\bibitem [{\citenamefont {Jendrzejewski}\ \emph {et~al.}(2014)\citenamefont
	{Jendrzejewski}, \citenamefont {Eckel}, \citenamefont {Murray}, \citenamefont
	{Lanier}, \citenamefont {Edwards}, \citenamefont {Lobb},\ and\ \citenamefont
	{Campbell}}]{ResistiveFlowRing}%
\BibitemOpen
\bibfield  {author} {\bibinfo {author} {\bibfnamefont {F.}~\bibnamefont
		{Jendrzejewski}}, \bibinfo {author} {\bibfnamefont {S.}~\bibnamefont
		{Eckel}}, \bibinfo {author} {\bibfnamefont {N.}~\bibnamefont {Murray}},
	\bibinfo {author} {\bibfnamefont {C.}~\bibnamefont {Lanier}}, \bibinfo
	{author} {\bibfnamefont {M.}~\bibnamefont {Edwards}}, \bibinfo {author}
	{\bibfnamefont {C.~J.}\ \bibnamefont {Lobb}}, \ and\ \bibinfo {author}
	{\bibfnamefont {G.~K.}\ \bibnamefont {Campbell}},\ }\href {\doibase
	10.1103/PhysRevLett.113.045305} {\bibfield  {journal} {\bibinfo  {journal}
		{Phys. Rev. Lett.}\ }\textbf {\bibinfo {volume} {113}},\ \bibinfo {pages}
	{045305} (\bibinfo {year} {2014})}\BibitemShut {NoStop}%
\bibitem [{\citenamefont {Eckel}\ \emph
	{et~al.}(2014{\natexlab{a}})\citenamefont {Eckel}, \citenamefont {Lee},
	\citenamefont {Jendrzejewski}, \citenamefont {Murray}, \citenamefont {Clark},
	\citenamefont {Lobb}, \citenamefont {Phillips}, \citenamefont {Edwards},\
	and\ \citenamefont {Campbell}}]{RingHysterisis}%
\BibitemOpen
\bibfield  {author} {\bibinfo {author} {\bibfnamefont {S.}~\bibnamefont
		{Eckel}}, \bibinfo {author} {\bibfnamefont {J.~G.}\ \bibnamefont {Lee}},
	\bibinfo {author} {\bibfnamefont {F.}~\bibnamefont {Jendrzejewski}}, \bibinfo
	{author} {\bibfnamefont {N.}~\bibnamefont {Murray}}, \bibinfo {author}
	{\bibfnamefont {C.~W.}\ \bibnamefont {Clark}}, \bibinfo {author}
	{\bibfnamefont {C.~J.}\ \bibnamefont {Lobb}}, \bibinfo {author}
	{\bibfnamefont {W.~D.}\ \bibnamefont {Phillips}}, \bibinfo {author}
	{\bibfnamefont {M.}~\bibnamefont {Edwards}}, \ and\ \bibinfo {author}
	{\bibfnamefont {G.~K.}\ \bibnamefont {Campbell}},\ }\href
{http://dx.doi.org/10.1038/nature12958} {\bibfield  {journal} {\bibinfo
		{journal} {Nature}\ }\textbf {\bibinfo {volume} {506}},\ \bibinfo {pages}
	{200} (\bibinfo {year} {2014}{\natexlab{a}})}\BibitemShut {NoStop}%
\bibitem [{\citenamefont {Aghamalyan}\ \emph {et~al.}(2015)\citenamefont
	{Aghamalyan}, \citenamefont {Cominotti}, \citenamefont {Rizzi}, \citenamefont
	{Rossini}, \citenamefont {Hekking}, \citenamefont {Minguzzi}, \citenamefont
	{Kwek},\ and\ \citenamefont {Amico}}]{RingAtomtronicInterference}%
\BibitemOpen
\bibfield  {author} {\bibinfo {author} {\bibfnamefont {D.}~\bibnamefont
		{Aghamalyan}}, \bibinfo {author} {\bibfnamefont {M.}~\bibnamefont
		{Cominotti}}, \bibinfo {author} {\bibfnamefont {M.}~\bibnamefont {Rizzi}},
	\bibinfo {author} {\bibfnamefont {D.}~\bibnamefont {Rossini}}, \bibinfo
	{author} {\bibfnamefont {F.}~\bibnamefont {Hekking}}, \bibinfo {author}
	{\bibfnamefont {A.}~\bibnamefont {Minguzzi}}, \bibinfo {author}
	{\bibfnamefont {L.-C.}\ \bibnamefont {Kwek}}, \ and\ \bibinfo {author}
	{\bibfnamefont {L.}~\bibnamefont {Amico}},\ }\href
{http://stacks.iop.org/1367-2630/17/i=4/a=045023} {\bibfield  {journal}
	{\bibinfo  {journal} {New Journal of Physics}\ }\textbf {\bibinfo {volume}
		{17}},\ \bibinfo {pages} {045023} (\bibinfo {year} {2015})}\BibitemShut
{NoStop}%
\bibitem [{\citenamefont {Wang}\ \emph {et~al.}(2015)\citenamefont {Wang},
	\citenamefont {Kumar}, \citenamefont {Jendrzejewski}, \citenamefont {Wilson},
	\citenamefont {Edwards}, \citenamefont {Eckel}, \citenamefont {Campbell},\
	and\ \citenamefont {Clark}}]{RingAtomtronicSquid}%
\BibitemOpen
\bibfield  {author} {\bibinfo {author} {\bibfnamefont {Y.-H.}\ \bibnamefont
		{Wang}}, \bibinfo {author} {\bibfnamefont {A.}~\bibnamefont {Kumar}},
	\bibinfo {author} {\bibfnamefont {F.}~\bibnamefont {Jendrzejewski}}, \bibinfo
	{author} {\bibfnamefont {R.~M.}\ \bibnamefont {Wilson}}, \bibinfo {author}
	{\bibfnamefont {M.}~\bibnamefont {Edwards}}, \bibinfo {author} {\bibfnamefont
		{S.}~\bibnamefont {Eckel}}, \bibinfo {author} {\bibfnamefont {G.~K.}\
		\bibnamefont {Campbell}}, \ and\ \bibinfo {author} {\bibfnamefont {C.~W.}\
		\bibnamefont {Clark}},\ }\href
{http://stacks.iop.org/1367-2630/17/i=12/a=125012} {\bibfield  {journal}
	{\bibinfo  {journal} {New Journal of Physics}\ }\textbf {\bibinfo {volume}
		{17}},\ \bibinfo {pages} {125012} (\bibinfo {year} {2015})}\BibitemShut
{NoStop}%
\bibitem [{\citenamefont {Mathey}\ and\ \citenamefont
	{Mathey}(2016)}]{AtomtronicSQUIDOptimization}%
\BibitemOpen
\bibfield  {author} {\bibinfo {author} {\bibfnamefont {A.~C.}\ \bibnamefont
		{Mathey}}\ and\ \bibinfo {author} {\bibfnamefont {L.}~\bibnamefont
		{Mathey}},\ }\href {http://stacks.iop.org/1367-2630/18/i=5/a=055016}
{\bibfield  {journal} {\bibinfo  {journal} {New Journal of Physics}\ }\textbf
	{\bibinfo {volume} {18}},\ \bibinfo {pages} {055016} (\bibinfo {year}
	{2016})}\BibitemShut {NoStop}%
\bibitem [{\citenamefont {Pepino}\ \emph {et~al.}(2009)\citenamefont {Pepino},
	\citenamefont {Cooper}, \citenamefont {Anderson},\ and\ \citenamefont
	{Holland}}]{AtomtronicCircuits}%
\BibitemOpen
\bibfield  {author} {\bibinfo {author} {\bibfnamefont {R.~A.}\ \bibnamefont
		{Pepino}}, \bibinfo {author} {\bibfnamefont {J.}~\bibnamefont {Cooper}},
	\bibinfo {author} {\bibfnamefont {D.~Z.}\ \bibnamefont {Anderson}}, \ and\
	\bibinfo {author} {\bibfnamefont {M.~J.}\ \bibnamefont {Holland}},\ }\href
{\doibase 10.1103/PhysRevLett.103.140405} {\bibfield  {journal} {\bibinfo
		{journal} {Phys. Rev. Lett.}\ }\textbf {\bibinfo {volume} {103}},\ \bibinfo
	{pages} {140405} (\bibinfo {year} {2009})}\BibitemShut {NoStop}%
\bibitem [{\citenamefont {Seaman}\ \emph {et~al.}(2007)\citenamefont {Seaman},
	\citenamefont {Kr\"amer}, \citenamefont {Anderson},\ and\ \citenamefont
	{Holland}}]{AtomtronicsAnalogs}%
\BibitemOpen
\bibfield  {author} {\bibinfo {author} {\bibfnamefont {B.~T.}\ \bibnamefont
		{Seaman}}, \bibinfo {author} {\bibfnamefont {M.}~\bibnamefont {Kr\"amer}},
	\bibinfo {author} {\bibfnamefont {D.~Z.}\ \bibnamefont {Anderson}}, \ and\
	\bibinfo {author} {\bibfnamefont {M.~J.}\ \bibnamefont {Holland}},\ }\href
{\doibase 10.1103/PhysRevA.75.023615} {\bibfield  {journal} {\bibinfo
		{journal} {Phys. Rev. A}\ }\textbf {\bibinfo {volume} {75}},\ \bibinfo
	{pages} {023615} (\bibinfo {year} {2007})}\BibitemShut {NoStop}%
\bibitem [{\citenamefont {Caliga}, \citenamefont {Straatsma},\ and\
	\citenamefont {Anderson}(2016)}]{AtomtronicTransistor}%
\BibitemOpen
\bibfield  {author} {\bibinfo {author} {\bibfnamefont {S.~C.}\ \bibnamefont
		{Caliga}}, \bibinfo {author} {\bibfnamefont {C.~J.~E.}\ \bibnamefont
		{Straatsma}}, \ and\ \bibinfo {author} {\bibfnamefont {D.~Z.}\ \bibnamefont
		{Anderson}},\ }\href {http://stacks.iop.org/1367-2630/18/i=2/a=025010}
{\bibfield  {journal} {\bibinfo  {journal} {New Journal of Physics}\ }\textbf
	{\bibinfo {volume} {18}},\ \bibinfo {pages} {025010} (\bibinfo {year}
	{2016})}\BibitemShut {NoStop}%
\bibitem [{\citenamefont {Caliga}\ \emph {et~al.}(2016)\citenamefont {Caliga},
	\citenamefont {Straatsma}, \citenamefont {Zozulya},\ and\ \citenamefont
	{Anderson}}]{AtomtronicTransistorPrinciple}%
\BibitemOpen
\bibfield  {author} {\bibinfo {author} {\bibfnamefont {S.~C.}\ \bibnamefont
		{Caliga}}, \bibinfo {author} {\bibfnamefont {C.~J.~E.}\ \bibnamefont
		{Straatsma}}, \bibinfo {author} {\bibfnamefont {A.~A.}\ \bibnamefont
		{Zozulya}}, \ and\ \bibinfo {author} {\bibfnamefont {D.~Z.}\ \bibnamefont
		{Anderson}},\ }\href {http://stacks.iop.org/1367-2630/18/i=1/a=015012}
{\bibfield  {journal} {\bibinfo  {journal} {New Journal of Physics}\ }\textbf
	{\bibinfo {volume} {18}},\ \bibinfo {pages} {015012} (\bibinfo {year}
	{2016})}\BibitemShut {NoStop}%
\bibitem [{\citenamefont {Lee}\ \emph {et~al.}(2013)\citenamefont {Lee},
	\citenamefont {McIlvain}, \citenamefont {Lobb},\ and\ \citenamefont
	{Hill}}]{ElectricAnalogsAtoms}%
\BibitemOpen
\bibfield  {author} {\bibinfo {author} {\bibfnamefont {J.~G.}\ \bibnamefont
		{Lee}}, \bibinfo {author} {\bibfnamefont {B.~J.}\ \bibnamefont {McIlvain}},
	\bibinfo {author} {\bibfnamefont {C.~J.}\ \bibnamefont {Lobb}}, \ and\
	\bibinfo {author} {\bibfnamefont {W.~T.}\ \bibnamefont {Hill}, \bibfnamefont
		{III}},\ }\href {\doibase 10.1038/srep01034} {\bibfield  {journal} {\bibinfo
		{journal} {Sci. Rep.}\ }\textbf {\bibinfo {volume} {3}},\ \bibinfo {pages}
	{1034} (\bibinfo {year} {2013})}\BibitemShut {NoStop}%
\bibitem [{\citenamefont {Eckel}\ \emph
	{et~al.}(2014{\natexlab{b}})\citenamefont {Eckel}, \citenamefont
	{Jendrzejewski}, \citenamefont {Kumar}, \citenamefont {Lobb},\ and\
	\citenamefont {Campbell}}]{WeakLink}%
\BibitemOpen
\bibfield  {author} {\bibinfo {author} {\bibfnamefont {S.}~\bibnamefont
		{Eckel}}, \bibinfo {author} {\bibfnamefont {F.}~\bibnamefont
		{Jendrzejewski}}, \bibinfo {author} {\bibfnamefont {A.}~\bibnamefont
		{Kumar}}, \bibinfo {author} {\bibfnamefont {C.~J.}\ \bibnamefont {Lobb}}, \
	and\ \bibinfo {author} {\bibfnamefont {G.~K.}\ \bibnamefont {Campbell}},\
}\href {\doibase 10.1103/PhysRevX.4.031052} {\bibfield  {journal} {\bibinfo
	{journal} {Phys. Rev. X}\ }\textbf {\bibinfo {volume} {4}},\ \bibinfo {pages}
{031052} (\bibinfo {year} {2014}{\natexlab{b}})}\BibitemShut {NoStop}%
\bibitem [{\citenamefont {Labouvie}\ \emph {et~al.}(2015)\citenamefont
	{Labouvie}, \citenamefont {Santra}, \citenamefont {Heun}, \citenamefont
	{Wimberger},\ and\ \citenamefont {Ott}}]{NegativeConductivity}%
\BibitemOpen
\bibfield  {author} {\bibinfo {author} {\bibfnamefont {R.}~\bibnamefont
		{Labouvie}}, \bibinfo {author} {\bibfnamefont {B.}~\bibnamefont {Santra}},
	\bibinfo {author} {\bibfnamefont {S.}~\bibnamefont {Heun}}, \bibinfo {author}
	{\bibfnamefont {S.}~\bibnamefont {Wimberger}}, \ and\ \bibinfo {author}
	{\bibfnamefont {H.}~\bibnamefont {Ott}},\ }\href {\doibase
	10.1103/PhysRevLett.115.050601} {\bibfield  {journal} {\bibinfo  {journal}
		{Phys. Rev. Lett.}\ }\textbf {\bibinfo {volume} {115}},\ \bibinfo {pages}
	{050601} (\bibinfo {year} {2015})}\BibitemShut {NoStop}%
\bibitem [{\citenamefont {Eckel}\ \emph {et~al.}(2016)\citenamefont {Eckel},
	\citenamefont {Lee}, \citenamefont {Jendrzejewski}, \citenamefont {Lobb},
	\citenamefont {Campbell},\ and\ \citenamefont {Hill}}]{Eckle}%
\BibitemOpen
\bibfield  {author} {\bibinfo {author} {\bibfnamefont {S.}~\bibnamefont
		{Eckel}}, \bibinfo {author} {\bibfnamefont {J.~G.}\ \bibnamefont {Lee}},
	\bibinfo {author} {\bibfnamefont {F.}~\bibnamefont {Jendrzejewski}}, \bibinfo
	{author} {\bibfnamefont {C.~J.}\ \bibnamefont {Lobb}}, \bibinfo {author}
	{\bibfnamefont {G.~K.}\ \bibnamefont {Campbell}}, \ and\ \bibinfo {author}
	{\bibfnamefont {W.~T.}\ \bibnamefont {Hill}},\ }\href {\doibase
	10.1103/PhysRevA.93.063619} {\bibfield  {journal} {\bibinfo  {journal} {Phys.
			Rev. A}\ }\textbf {\bibinfo {volume} {93}},\ \bibinfo {pages} {063619}
	(\bibinfo {year} {2016})}\BibitemShut {NoStop}%
\bibitem [{\citenamefont {Abrahams}(2010)}]{50_years_anderson}%
\BibitemOpen
\bibfield  {author} {\bibinfo {author} {\bibfnamefont {E.}~\bibnamefont
		{Abrahams}},\ }\href@noop {} {\emph {\bibinfo {title} {50 Years of Anderson
			Localization}}}\ (\bibinfo {year} {2010})\BibitemShut {NoStop}%
\bibitem [{\citenamefont {Gooch}\ and\ \citenamefont
	{Tarry}(1975)}]{liquid_crystal}%
\BibitemOpen
\bibfield  {author} {\bibinfo {author} {\bibfnamefont {C.~H.}\ \bibnamefont
		{Gooch}}\ and\ \bibinfo {author} {\bibfnamefont {H.~A.}\ \bibnamefont
		{Tarry}},\ }\href {http://stacks.iop.org/0022-3727/8/i=13/a=020} {\bibfield
	{journal} {\bibinfo  {journal} {Journal of Physics D: Applied Physics}\
	}\textbf {\bibinfo {volume} {8}},\ \bibinfo {pages} {1575} (\bibinfo {year}
	{1975})}\BibitemShut {NoStop}%
\bibitem [{\citenamefont {Gerchberg}\ and\ \citenamefont
	{Saxtion}(1972)}]{GS_algorithm}%
\BibitemOpen
\bibfield  {author} {\bibinfo {author} {\bibfnamefont {R.~W.}\ \bibnamefont
		{Gerchberg}}\ and\ \bibinfo {author} {\bibfnamefont {W.~O.}\ \bibnamefont
		{Saxtion}},\ }\href {http://ci.nii.ac.jp/naid/10010556614/en/} {\bibfield
	{journal} {\bibinfo  {journal} {Optik}\ }\textbf {\bibinfo {volume} {35}},\
	\bibinfo {pages} {237} (\bibinfo {year} {1972})}\BibitemShut {NoStop}%
\bibitem [{\citenamefont {Pasienski}\ and\ \citenamefont
	{DeMarco}(2008)}]{mraf}%
\BibitemOpen
\bibfield  {author} {\bibinfo {author} {\bibfnamefont {M.}~\bibnamefont
		{Pasienski}}\ and\ \bibinfo {author} {\bibfnamefont {B.}~\bibnamefont
		{DeMarco}},\ }\href {\doibase 10.1364/OE.16.002176} {\bibfield  {journal}
	{\bibinfo  {journal} {Opt. Express}\ }\textbf {\bibinfo {volume} {16}},\
	\bibinfo {pages} {2176} (\bibinfo {year} {2008})}\BibitemShut {NoStop}%
\bibitem [{\citenamefont {Gaunt}\ and\ \citenamefont
	{Hadzibabic}(2012)}]{omraf}%
\BibitemOpen
\bibfield  {author} {\bibinfo {author} {\bibfnamefont {A.~L.}\ \bibnamefont
		{Gaunt}}\ and\ \bibinfo {author} {\bibfnamefont {Z.}~\bibnamefont
		{Hadzibabic}},\ }\href {http://dx.doi.org/10.1038/srep00721} {\bibfield
	{journal} {\bibinfo  {journal} {Scientific Reports}\ }\textbf {\bibinfo
		{volume} {2}},\ \bibinfo {pages} {721} (\bibinfo {year} {2012})}\BibitemShut
{NoStop}%
\bibitem [{\citenamefont {Garc\'{i}a-M\'{a}rquez}\ \emph
	{et~al.}(2012)\citenamefont {Garc\'{i}a-M\'{a}rquez}, \citenamefont
	{L\'{o}pez}, \citenamefont {Gonz\'{a}lez-Vega},\ and\ \citenamefont
	{No\'{e}}}]{Garcia-Marquez:12}%
\BibitemOpen
\bibfield  {author} {\bibinfo {author} {\bibfnamefont {J.}~\bibnamefont
		{Garc\'{i}a-M\'{a}rquez}}, \bibinfo {author} {\bibfnamefont {V.}~\bibnamefont
		{L\'{o}pez}}, \bibinfo {author} {\bibfnamefont {A.}~\bibnamefont
		{Gonz\'{a}lez-Vega}}, \ and\ \bibinfo {author} {\bibfnamefont
		{E.}~\bibnamefont {No\'{e}}},\ }\href {\doibase 10.1364/OE.20.008431}
{\bibfield  {journal} {\bibinfo  {journal} {Opt. Express}\ }\textbf {\bibinfo
		{volume} {20}},\ \bibinfo {pages} {8431} (\bibinfo {year}
	{2012})}\BibitemShut {NoStop}%
\bibitem [{\citenamefont {Jo}\ \emph {et~al.}(2012)\citenamefont {Jo},
	\citenamefont {Guzman}, \citenamefont {Thomas}, \citenamefont {Hosur},
	\citenamefont {Vishwanath},\ and\ \citenamefont {Stamper-Kurn}}]{Kagome}%
\BibitemOpen
\bibfield  {author} {\bibinfo {author} {\bibfnamefont {G.-B.}\ \bibnamefont
		{Jo}}, \bibinfo {author} {\bibfnamefont {J.}~\bibnamefont {Guzman}}, \bibinfo
	{author} {\bibfnamefont {C.~K.}\ \bibnamefont {Thomas}}, \bibinfo {author}
	{\bibfnamefont {P.}~\bibnamefont {Hosur}}, \bibinfo {author} {\bibfnamefont
		{A.}~\bibnamefont {Vishwanath}}, \ and\ \bibinfo {author} {\bibfnamefont
		{D.~M.}\ \bibnamefont {Stamper-Kurn}},\ }\href {\doibase
	10.1103/PhysRevLett.108.045305} {\bibfield  {journal} {\bibinfo  {journal}
		{Phys. Rev. Lett.}\ }\textbf {\bibinfo {volume} {108}},\ \bibinfo {pages}
	{045305} (\bibinfo {year} {2012})}\BibitemShut {NoStop}%
\bibitem [{\citenamefont {Tarruell}\ \emph {et~al.}(2012)\citenamefont
	{Tarruell}, \citenamefont {Greif}, \citenamefont {Uehlinger}, \citenamefont
	{Jotzu},\ and\ \citenamefont {Esslinger}}]{honeycomb}%
\BibitemOpen
\bibfield  {author} {\bibinfo {author} {\bibfnamefont {L.}~\bibnamefont
		{Tarruell}}, \bibinfo {author} {\bibfnamefont {D.}~\bibnamefont {Greif}},
	\bibinfo {author} {\bibfnamefont {T.}~\bibnamefont {Uehlinger}}, \bibinfo
	{author} {\bibfnamefont {G.}~\bibnamefont {Jotzu}}, \ and\ \bibinfo {author}
	{\bibfnamefont {T.}~\bibnamefont {Esslinger}},\ }\href
{http://dx.doi.org/10.1038/nature10871} {\bibfield  {journal} {\bibinfo
		{journal} {Nature}\ }\textbf {\bibinfo {volume} {483}},\ \bibinfo {pages}
	{302} (\bibinfo {year} {2012})}\BibitemShut {NoStop}%
\bibitem [{\citenamefont {Robens}\ \emph {et~al.}(2017)\citenamefont {Robens},
	\citenamefont {Brakhane}, \citenamefont {Alt}, \citenamefont {Klei{\ss}ler},
	\citenamefont {Meschede}, \citenamefont {Moon}, \citenamefont {Ramola},\ and\
	\citenamefont {Alberti}}]{Robens:17}%
\BibitemOpen
\bibfield  {author} {\bibinfo {author} {\bibfnamefont {C.}~\bibnamefont
		{Robens}}, \bibinfo {author} {\bibfnamefont {S.}~\bibnamefont {Brakhane}},
	\bibinfo {author} {\bibfnamefont {W.}~\bibnamefont {Alt}}, \bibinfo {author}
	{\bibfnamefont {F.}~\bibnamefont {Klei{\ss}ler}}, \bibinfo {author}
	{\bibfnamefont {D.}~\bibnamefont {Meschede}}, \bibinfo {author}
	{\bibfnamefont {G.}~\bibnamefont {Moon}}, \bibinfo {author} {\bibfnamefont
		{G.}~\bibnamefont {Ramola}}, \ and\ \bibinfo {author} {\bibfnamefont
		{A.}~\bibnamefont {Alberti}},\ }\href {\doibase 10.1364/OL.42.001043}
{\bibfield  {journal} {\bibinfo  {journal} {Opt. Lett.}\ }\textbf {\bibinfo
		{volume} {42}},\ \bibinfo {pages} {1043} (\bibinfo {year}
	{2017})}\BibitemShut {NoStop}%
\bibitem [{\citenamefont {Leung}\ \emph {et~al.}(2014)\citenamefont {Leung},
	\citenamefont {Pijn}, \citenamefont {Schlatter}, \citenamefont
	{Torralbo-Campo}, \citenamefont {Rooij}, \citenamefont {Mulder},
	\citenamefont {Naber}, \citenamefont {Soudijn}, \citenamefont {Tauschinsky},
	\citenamefont {Abarbanel}, \citenamefont {Hadad}, \citenamefont {Golan},
	\citenamefont {Folman},\ and\ \citenamefont {Spreeuw}}]{Leung:14}%
\BibitemOpen
\bibfield  {author} {\bibinfo {author} {\bibfnamefont {V.~Y.~F.}\
		\bibnamefont {Leung}}, \bibinfo {author} {\bibfnamefont {D.~R.~M.}\
		\bibnamefont {Pijn}}, \bibinfo {author} {\bibfnamefont {H.}~\bibnamefont
		{Schlatter}}, \bibinfo {author} {\bibfnamefont {L.}~\bibnamefont
		{Torralbo-Campo}}, \bibinfo {author} {\bibfnamefont {A.~L.~L.}\ \bibnamefont
		{Rooij}}, \bibinfo {author} {\bibfnamefont {G.~B.}\ \bibnamefont {Mulder}},
	\bibinfo {author} {\bibfnamefont {J.}~\bibnamefont {Naber}}, \bibinfo
	{author} {\bibfnamefont {M.~L.}\ \bibnamefont {Soudijn}}, \bibinfo {author}
	{\bibfnamefont {A.}~\bibnamefont {Tauschinsky}}, \bibinfo {author}
	{\bibfnamefont {C.}~\bibnamefont {Abarbanel}}, \bibinfo {author}
	{\bibfnamefont {B.}~\bibnamefont {Hadad}}, \bibinfo {author} {\bibfnamefont
		{E.}~\bibnamefont {Golan}}, \bibinfo {author} {\bibfnamefont
		{R.}~\bibnamefont {Folman}}, \ and\ \bibinfo {author} {\bibfnamefont
		{R.~J.~C.}\ \bibnamefont {Spreeuw}},\ }\href {\doibase 10.1063/1.4874005}
{\bibfield  {journal} {\bibinfo  {journal} {Review of Scientific
			Instruments}\ }\textbf {\bibinfo {volume} {85}},\ \bibinfo {pages} {053102}
	(\bibinfo {year} {2014})},\ \Eprint
{http://arxiv.org/abs/http://dx.doi.org/10.1063/1.4874005}
{http://dx.doi.org/10.1063/1.4874005} \BibitemShut {NoStop}%
\bibitem [{\citenamefont {Hecht}(2017)}]{Hecht2017}%
\BibitemOpen
\bibfield  {author} {\bibinfo {author} {\bibfnamefont {E.}~\bibnamefont
		{Hecht}},\ }\href@noop {} {\emph {\bibinfo {title} {Optics}}},\ \bibinfo
{edition} {5th}\ ed.\ (\bibinfo  {publisher} {Harlow: Pearson Education},\
\bibinfo {year} {2017})\BibitemShut {NoStop}%
\bibitem [{\citenamefont {Wenas}\ and\ \citenamefont
	{Hoogerland}(2008)}]{Wenas}%
\BibitemOpen
\bibfield  {author} {\bibinfo {author} {\bibfnamefont {Y.~C.}\ \bibnamefont
		{Wenas}}\ and\ \bibinfo {author} {\bibfnamefont {M.~D.}\ \bibnamefont
		{Hoogerland}},\ }\href {\doibase 10.1063/1.2917405} {\bibfield  {journal}
	{\bibinfo  {journal} {Rev. Sci. Instrum.}\ }\textbf {\bibinfo {volume}
		{79}},\ \bibinfo {pages} {053101} (\bibinfo {year} {2008})}\BibitemShut
{NoStop}%
\bibitem [{\citenamefont {Li}\ \emph {et~al.}(2016)\citenamefont {Li},
	\citenamefont {Eckel}, \citenamefont {Eller}, \citenamefont {Warren},
	\citenamefont {Clark},\ and\ \citenamefont {Edwards}}]{PhysRevA.94.023626}%
\BibitemOpen
\bibfield  {author} {\bibinfo {author} {\bibfnamefont {A.}~\bibnamefont
		{Li}}, \bibinfo {author} {\bibfnamefont {S.}~\bibnamefont {Eckel}}, \bibinfo
	{author} {\bibfnamefont {B.}~\bibnamefont {Eller}}, \bibinfo {author}
	{\bibfnamefont {K.~E.}\ \bibnamefont {Warren}}, \bibinfo {author}
	{\bibfnamefont {C.~W.}\ \bibnamefont {Clark}}, \ and\ \bibinfo {author}
	{\bibfnamefont {M.}~\bibnamefont {Edwards}},\ }\href {\doibase
	10.1103/PhysRevA.94.023626} {\bibfield  {journal} {\bibinfo  {journal} {Phys.
			Rev. A}\ }\textbf {\bibinfo {volume} {94}},\ \bibinfo {pages} {023626}
	(\bibinfo {year} {2016})}\BibitemShut {NoStop}%
\bibitem [{\citenamefont {Bruce}\ \emph {et~al.}(2011)\citenamefont {Bruce},
	\citenamefont {Mayoh}, \citenamefont {Smirne}, \citenamefont
	{Torralbo-Campo},\ and\ \citenamefont
	{Cassettari}}]{ring_trap_Cassettari_group}%
\BibitemOpen
\bibfield  {author} {\bibinfo {author} {\bibfnamefont {G.~D.}\ \bibnamefont
		{Bruce}}, \bibinfo {author} {\bibfnamefont {J.}~\bibnamefont {Mayoh}},
	\bibinfo {author} {\bibfnamefont {G.}~\bibnamefont {Smirne}}, \bibinfo
	{author} {\bibfnamefont {L.}~\bibnamefont {Torralbo-Campo}}, \ and\ \bibinfo
	{author} {\bibfnamefont {D.}~\bibnamefont {Cassettari}},\ }\href
{http://stacks.iop.org/1402-4896/2011/i=T143/a=014008} {\bibfield  {journal}
	{\bibinfo  {journal} {Physica Scripta}\ }\textbf {\bibinfo {volume} {2011}},\
	\bibinfo {pages} {014008} (\bibinfo {year} {2011})}\BibitemShut {NoStop}%
\end{thebibliography}

%
\end{document}